\documentstyle[12pt]{article}
\newcommand{\lettersize}{\baselineskip=0.8cm}
\newcommand{\dir}{FIGURES}
\input psfig
\input{psfig}
\newcommand{\fig}[3]
{
\begin{center}
     \noindent
     \unitlength=1mm
     \begin{picture}(#2,#3)
     \put(0,0){
       \psfig{figure=\dir/#1,width=#2mm,height=#3mm}
     }
     \end{picture}
   \noindent
\end{center}
}
\begin{document}
\baselineskip=12pt
\setcounter{page}{1}

\begin{center}
{\LARGE\bf
Systems involving surfactants
}
\end{center}

\vspace{0.5cm}

\lettersize

\begin{center}
Friederike Schmid \\
{\em Institut f\"ur Physik, Universit\"at Mainz, D-55099 Mainz, FRG}
\end{center}

\begin{quote}
{\bf Abstract.}
Computer simulations of amphiphilic systems are reviewed. 
Research areas cover a wide range of length and time scales, and a whole 
hierarchy of models and methods has been developed to address them all. 
They range from atomistically realistic models,
idealized chain models, lattice spin models, to phenomenological 
models such as Ginzburg-Landau models and random interface models. 
Selected applications are discussed in order to illustrate the
use of the models and the insights they can offer. 
\end{quote}

\vspace{2cm}


\lettersize

\section{Introduction}
\label{s1}

\begin{figure}
\fig{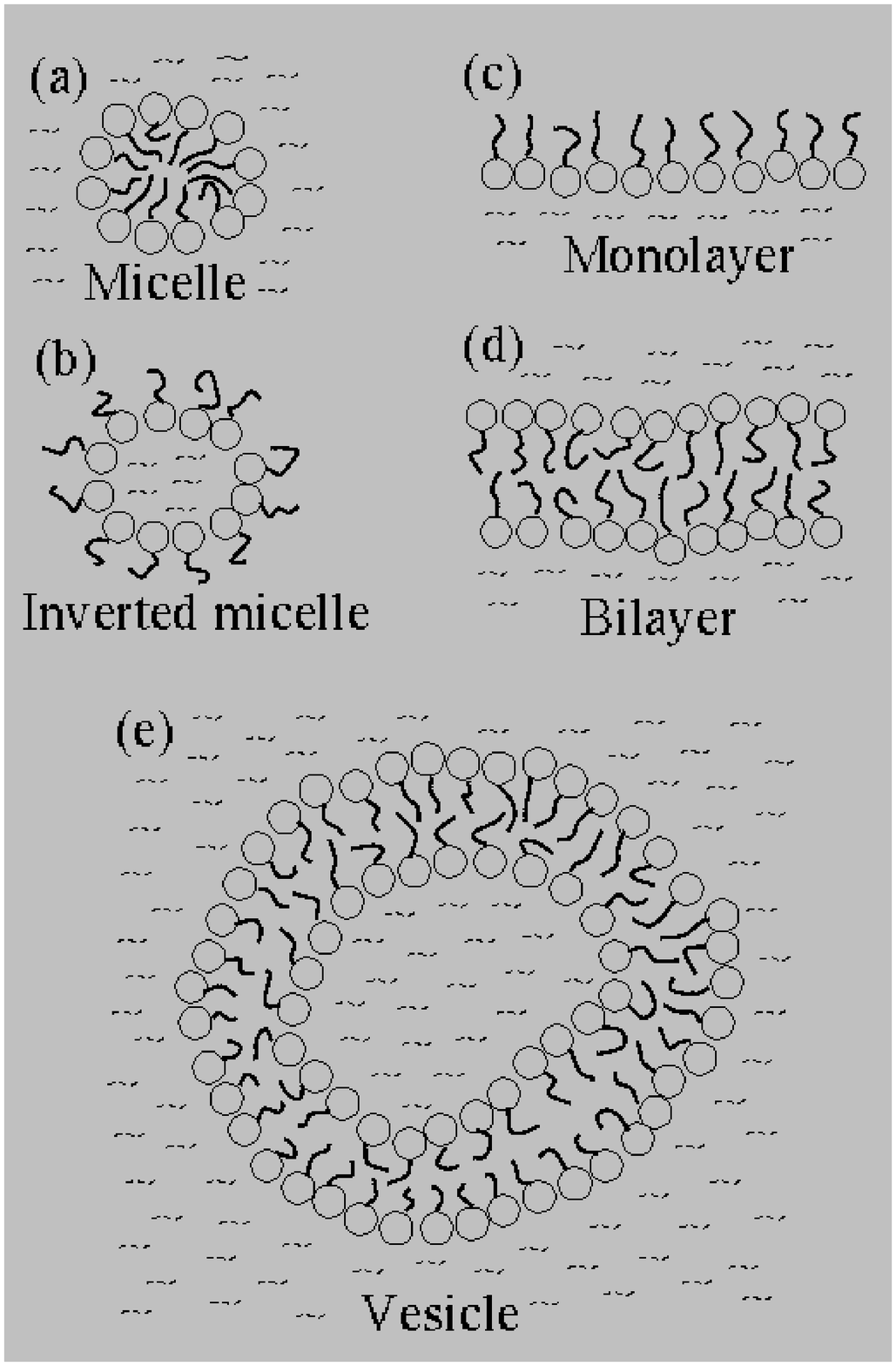}{80}{100}
\caption{\label{fig1}}
\end{figure}

Surfactants can be defined very generally as substances which influence the 
properties of interfaces and surfaces, and which can be used to tune 
them. Since most materials contain a certain amount of internal interfaces, 
the study of such substances has attracted longstanding interest.
In this chapter, we shall be concerned with a particularly
efficient class of surfactants, the amphiphiles\cite{gs_review,amph_books}. 
These are molecules which contain both hydrophilic and hydrophobic units 
(usually one or several hydrocarbon chains), such that they
``love'' and ``hate'' water at the same time. Familiar examples are lipids 
or alcohols. The effect of amphiphiles on interfaces between water and 
nonpolar phases can be quite dramatic. For example, tiny additions of 
good amphiphiles reduce the interfacial tension by several orders of 
magnitude. Amphiphiles are thus very efficient to promote the dispersion of
organic fluids in water and vice versa. Added in larger amounts, they 
associate into a variety of structures, filling the material with internal 
interfaces, which shield the oil molecules -- or in the absence of oil the 
hydrophobic parts of the amphiphiles -- from the water\cite{amph_exp}. 
Some of the possible structures are depicted in Fig. \ref{fig1}. 
A very rich phase behavior emerges, including isotropic, but mesoscopically 
structured micellar and bicontinuous phases, and several ordered anisotropic 
phases (Fig. \ref{fig2}). 
Due to their emulsifying and self-organizing qualities, amphiphilic molecules 
are widely used in technology (salad dressing, detergents, soaps, 
oil recovery, as coating materials to stabilize colloidal systems, 
as nanoreactors for the preparation of nanoparticles,...)
as well as by nature (milk, biological membranes, liposomes,...).

Schematic phase diagrams for binary mixtures of water with a strong 
amphiphile, and for ternary mixtures containing oil, water, and amphiphile, 
are shown in Fig. \ref{fig3} (adapted from Refs. \cite{strey1,strey2}). 
Among the many interesting features of these phase diagrams, we
mention in particular the existence of phases which are macroscopically
isotropic and homogeneous, but structured on a mesoscopic scale. 
The short range order can be inferred from the form of the structure 
functions (obtained, {\em e.g.}, by small angle neutron scattering). In the
case of microemulsions, the best indicator for structure is the
water-water structure function. It can be fitted very well by the 
expression\cite{strey3}
\begin{equation}
\label{s}
 S(q) \propto (a+g q^2 + c q^4)^{-1}
\end{equation}
For strongly structured microemulsions, $g$ is negative, and the structure 
functions show a peak at nonzero wavevector $q$. As long as 
$g < 2 \sqrt{c a}$, inverse Fourier transform of $S(q)$ still reveals 
that the water-water correlation functions oscillate rather than
decay monotonically. The lines in phase space where this oscillating behavior 
sets in are usually referred to as ``disorder lines'', and those where the 
maximum of $S(q)$ moves away from zero as ``Lifshitz lines''.

Another phase which has attracted recent interest is the gyroid phase, 
a bicontinuous ordered phase with cubic symmetry 
(space group Ia$\bar{3}$d, cf. Fig. \ref{fig2} (d)\cite{michael_gyroid}. 
It consists of two interwoven, but unconnected bicontinuous networks.
The amphiphile sheets have a mean curvature which is close to constant and 
intermediate between that of the usually neighboring lamellar and hexagonal 
phases. The gyroid phase has first been identified in lipid/water 
mixtures\cite{lipid_gyroid}, and has been found in many related systems 
since then, among other in copolymer blends\cite{copolymer_gyroid}.

In a somewhat wider sense, one can define amphiphiles as molecules
in which chemically very different units are linked together. 
For example, the structures formed by A:B block copolymers in demixed
A and/or B homopolymer melts and their phase behavior is very similar
to that of classical amphiphiles in water and/or oil\cite{copph,hasegawa}.
Copolymers are used to disperse immiscible homopolymer phases in
one another, but also to create new, mesoscopically structured
materials with unusual and interesting properties\cite{templin}.

Another interesting class of phase transitions are internal transitions 
within amphiphilic monolayers or bilayers. In particular, monolayers of
amphiphiles at the air/water interface (Langmuir monolayers) have been 
intensively studied in the past as experimentally fairly accessible 
model systems\cite{mono_reviews,kaganer}. A schematic phase diagram
for long chain fatty acids, alcohols, or lipids is shown in Fig. \ref{fig4}. 
On increasing the area per molecule, one observes two distinct coexistence 
regions between fluid phases: A transition from a highly diluted, 
``gas''like phase into a more condensed, ``liquid expanded'' phase, and a 
second transition into an even denser ``liquid condensed'' region.
The latter splits into a number of distinct phases, which differ in their
tilt order, positional order and orientational order of the backbones of the 
(CH${}_2$)${}_n$ chains. In bilayers, the liquid expanded/liquid condensed
transition translates into the ``main transition'' or the ``liquid/gel'' 
transition, where the bilayer thickness, the amphiphile mobility, and the 
conformational order of the chains, jump discontinuously as a function of 
the temperature\cite{bilayer_reviews}. On the ordered gel side various 
different phases exist, among them tilted phases and wavy phases 
(ripple phases). Bilayer phases are of special interest from a biological 
point of view, since the structure of biological membranes is coupled to 
the membrane functions\cite{bilayer_reviews}.

This rich scenario and the wide range of applications of amphiphiles
is reflected by an equally wide range of problems which have been
addressed in studies of amphiphilic systems. In particular, the interest 
has focussed on the following rather different classes of topics:
\begin{itemize}
\item The influence of amphiphiles on interfacial properties:
  Interfacial tension, wetting behavior, dynamical aspects such as the 
  question how small amounts of surfactant influence the kinetics of 
  phase separation.
\item The reasons for self-assembly and the mechanisms:
  Necessary conditions for the aggregation into micelles, mono- or bilayers,
  structure of aggregates, distribution of aggregation numbers etc.
\item Mesoscopic structures and phases:
  vesicles and vesicle shapes, structured phases and phase behavior 
  of amphiphilic systems.
\item The inner structure of monolayers and bilayers,
  the liquid/gel transition, tilt transitions, mixed layers etc. 
\end{itemize}  

A wide range of length scales and time scales is involved in these different
problems, and a unified treatment of amphiphilic systems is practically 
impossible. On the smallest scales (up to $\le$ 100 \AA \quad and 
10-100 ns\cite{fn1}), molecular dynamics simulations in atomic detail are 
feasible\cite{penna,kleinr}. They provide information on the microscopic 
structure and the short-time dynamics of specific systems and have the big 
advantage that they allow for a direct, quantitative comparison with 
experiments. On the other hand, they rely heavily on the quality of the 
force fields. The choice of the model is the most important and
the most critical step in an atomistic simulation. Other limitations
come from the short accessible time scales. Phases must usually be 
preassembled since the times for self-assembly are very large. 
Hence it is difficult to decide whether a given structure is metastable 
or really corresponds to a true free energy minimum. Mapping out a phase 
diagram by a systematic variation of parameters (temperature, pressure) is
generally out of reach. However, the force fields are improving, the computer 
technology is developing, and realistic simulations of amphiphilic
systems can be pushed further and further. A large number of studies 
has dealt with the local properties of micelles, monolayers, and particularly 
of lipid bilayers. This is a rapidly growing field of research, 
and enormous progress has been achieved in the last few years. 
A brief overview over some of the activities shall be given in 
section \ref{s2}.

Nevertheless, large scale phenomena and complicated phase diagrams 
cannot be investigated within realistic models at the moment, and
this is not very likely to change soon.
Therefore, theorists have often resorted to coarse-grained models, which
capture the features of the substances believed to be essential for the
properties of interest. Such models can provide qualitative and
semiquantitative insight into the physics of these materials,
and hopefully establish general relationships between microscopic and 
thermodynamic quantities. 

What are the essential features of surfactant systems?
An important ingredient is obviously the repulsion between water and
nonpolar molecules or molecule parts, the hydrophobic force.
This interaction is however highly nontrivial, and its analysis is still
an active field of research\cite{israelachvili,tanford,hydrophobic}. 
Qualitatively, it is usually attributed to the strong orientational and 
positional correlations between nonpolar molecules in solution and the 
surrounding water molecules. The origin of the interaction is therefore
entropic: Free water forms a network of hydrogen bonds. In the neighborhood 
of a nonpolar solute, the water molecules reorient and replace themselves 
so as to maintain the highest possible number of hydrogen bonds. The
associated loss of entropy accounts for the dominant contribution to the
hydrophobic interaction.
This effect is of course far too involved to be easily included in
a simplified model for amphiphilic systems. On the other hand,
we have already noted that many characteristica of amphiphilic systems
are also found in homopolymer/copolymer mixtures, where the interactions 
can be of very different nature. Hence the precise origin of the interactions
does not seem to be important for the phenomenology, and they can
conceivably be replaced by simpler potentials for most purposes. 

The second important attribute of amphiphiles is their affinity to
both water and oil. This aspect is retained in the microscopic models,
which will be discussed in sections \ref{s3} and \ref{s4}.
Oil, water and surfactant molecules are represented by simplified 
pseudoparticles. 

In some of these models (see section \ref{s3}), the surfactants are still 
treated as flexible chains\cite{tanny}. This allows one to study the role of 
the chain length and chain conformations. For example, the chain degrees of 
freedom are responsible for the internal phase transitions in monolayers and 
bilayers, in particular the liquid/gel transition. The chain length and chain
architecture determines the efficiency of an amphiphile and thus influences
the phase behavior. Moreover, it affects the shapes and size distributions
of micelles. Chain models are usually fairly universal, in the sense that 
they can be used to study  many different phenomena.

Models of a second type (section \ref{s4}) restrict themselves to a few very 
basic ingredients, {\em e.g.}, the repulsion between oil and water and the 
orientation of the amphiphiles. They are less versatile than chain models and 
have to be specified in view of the particular problem one has in mind. On the
other hand, they allow for an efficient study of structures on intermediate 
length and time scales, while still establishing a connection with microscopic 
properties of the materials. Hence, they bridge between the microscopic 
approaches and the more phenomenological treatments which will be described 
below. Various microscopic models of this type have been constructed and used 
to study phase transitions in the bulk of amphiphilic systems, internal phase 
transitions in monolayers and bilayers, interfacial properties, and 
dynamical aspects such as the kinetics of phase separation between water
and oil at the presence of amphiphiles.

Finally, when it comes to large scale structures and long time dynamics, 
it often proves useful to drop the notion of particles altogether
and to resort to phenomenological models (see section \ref{s5}).
Here again, one can distinguish between two main lines of approaches.
Ginzburg-Landau theories\cite{gs_review,gerhardr1} characterize the system by 
smooth ``order parameters'', which stand for local, coarse-grained averages of 
microscopic quantities. The coarse graining length is usually not specified, 
but thought to be of the order of the bulk correlation length, {\em i.e.}, a 
few molecular diameters. The order parameter fields are then distributed 
according to a free energy functional, which is typically constructed
from symmetry considerations as an expansion in powers of the order parameter.
Generic behavior is obtained as a function of the coefficients in the
expansion. Unfortunately, the relation between the model parameters and the 
``real'' parameters such as the pressure and the temperature is usually
not obvious. On the other hand, Ginzburg-Landau theories have the advantage 
that they require relatively little input, which can in part be taken 
from experiments ({\em e.g.}, from scattering experiments). 
Like the simple microscopic models, they have been applied to the study of 
phase behavior, interfaces, and the phase separation kinetics in 
amphiphilic mixtures. We will very briefly sketch the approach in
section \ref{s51} and refer to chapter 11 for a more detailed discussion.

The other class of phenomenological approaches subsumes the random surface 
theories (section \ref{s52}). These reduce the system to a set of internal
surfaces, supposedly filled with amphiphiles, which can be described by an 
effective interface Hamiltonian. The internal surfaces represent either 
bilayers or monolayers -- bilayers in binary amphiphile-water mixtures, and 
monolayers in ternary mixtures, where the monolayers are assumed to separate 
oil domains from water domains. Random surface theories have been formulated 
on lattices and in the continuum. In the latter case, they are an interesting 
application of the membrane theories, which are studied in many areas of 
physics, from general statistical field theory to elementary particle 
physics\cite{nelson_book}. Random surface theories for amphiphilic systems 
have been used to calculate shapes and distributions of vesicles, and phase 
transitions\cite{safran_book,peliti1,udo,gerhardr2,mouritsenr1}.

We close these introductory remarks with a few comments on the methods
which are actually used to study these models. They shall for the most
part only be mentioned very briefly. In the rest of this chapter,
we will focus mainly on computer simulations. Even those shall not
be explained in detail, for the simple reason that the models are too
different and the simulation methods too many. Rather, we refer the reader 
to the available textbooks on simulation methods, {\em e.g.}, 
Ref. \cite{bh,at,binder1,como}, and discuss only a few technical aspects here.  
In the case of atomistically realistic models, simulations are indeed the 
only possible way to approach these systems. Idealized microscopic models have 
usually been explored extensively by mean field methods. Even those can 
become quite involved for complex models, especially for chain models. 
One particularly popular and successful method to deal with chain molecules
has been the self-consistent field theory. In a nutshell, it treats
chains as random walks in a positional dependent chemical potential, which 
depends in turn on the conformational distributions of the chains in some
self-consistent way. A recent survey of the method can be found in 
Ref. \cite{friederike1}.
Self-consistent field approaches provide extremely good descriptions of 
polymer blends (for which they have been developed originally), since 
macromolecules have many contacts with other molecules and therefore show
mean-field type behavior almost everywhere in phase space.
In the case of short chain surfactants in low molecular weight solvents, 
the quantitative predictions are not quite as satisfactory. In particular, 
the stability of self-assembled aggregates can be overestimated by several 
orders of magnitude\cite{chris1}. However, the local structure in the 
aggregates, local density profiles etc., are still in reasonable agreement 
with Monte Carlo simulations\cite{chris1,benshaul2}.
Self-consistent field theories have been used to study 
micelles\cite{chris1,szleifer}, bilayers\cite{szleifer,benshaul1,benshaul2} 
and monolayers\cite{leermakers,friederike2,rieu}. More details on the
technique are given in chapter 9 of this book.

The first step to study phenomenological theories (Ginzburg-Landau theories
and membrane theories) has usually been to minimize the free energy 
functional of the model. Fluctuations are then included at a later stage, 
{\em e.g.}, using Monte Carlo simulations. The latter shall be 
discussed in section \ref{s5} and in chapter 11.

\section{Simulations in atomic detail}
\label{s2}

Even among the so-called realistic simulations, there are still very
different levels of coarse-graining. A full quantum chemical and ab-initio 
treatment of a system as complex as amphiphilic systems is practically 
impossible. Hence they are usually studied by classical molecular dynamics 
in combination with empirical interaction potentials, which have been 
optimized by fitting them to experimental data (structural data, latent heat 
of first order transitions etc.) and/or to ab-initio data (substructures of 
molecules). The functional form of these potentials may have little or nothing
to do with the underlying physics of the interactions. For example, 
the forces between atoms of different chains are usually taken to be 
pairwise additive, even if the underlying interactions are nonadditive 
dispersion forces\cite{israelachvili}. 
One has thus some freedom in the choice of the potentials, and many different 
forms are available. Hydrogen atoms are sometimes not included explicitly, but
adsorbed into the more ``important'' neighbor atoms (united atom models vs. 
full atomic detail).  Bond angles and bond lengths are sometimes constrained 
to take fixed values (constrained models vs. fully flexible models).
Complicated head groups are sometimes replaced by simpler structures. 
The models also differ in the way charges and partial charges are treated. 
Some models do not include electrostatic interactions at all, {\em i.e.}, 
they have no $1/r$ terms in the potentials. Others do, but truncate them 
at a certain distance, and still others deal with them in full by Ewald 
summation methods\cite{ewald}. Whether or not interactions are truncated 
obviously makes a difference\cite{feller3}. 
One could hope that the long range character of the interactions is not 
essential for the physics of amphiphilic systems, in which case one empirical 
potential is as promising as the other, as long as they all have been 
optimized independently. Note that long range interactions are ignored in 
most of the more idealized treatments of amphiphilic systems. On the other
hand, a careful treatment of the electrostatic interactions is crucial to 
reproduce the peculiarities of water, such as the dielectric 
properties\cite{hunenberger}. 

After these introductory {\em caveats}, we review some of the work
that has been done in realistic simulations of amphiphilic systems.

Perhaps the first simulation of a system with amphiphiles is due to Kox
in 1980\cite{kox}, a model of a lipid monolayer. A number of simulations of 
monolayers at air/substrate interfaces have followed\cite{northrup,alper,
klein1,harris1,clough,moller,ahlstroem,karaborni1,siepmann},
which have been able to reproduce cooperative tilt effects of 
the hydrocarbon chains and tilt transitions as well as backbone ordering 
and different rotator phases\cite{klein1,karaborni1}. Moreover, they have
provided valuable insight on the density profiles and the distribution
of defects in the chains\cite{moller,karaborni1}. 
The simulations are usually done at constant volume. In one case,
the Gibbs ensemble is used to study the phase coexistence between 
the ``gas'' phase and the ``liquid expanded'' phase\cite{siepmann}
(In the Gibbs ensemble, two boxes are simulated in parallel, and 
particles can be exchanged between the boxes\cite{smit_r1}.
The two boxes are thus at pressure and chemical potential equilibrium:
Coexisting phases can be monitored in systems which contain
no disturbing interfaces.)
In most of the models, the amphiphiles are supported by a uniform continuum 
substrate, which could be water or any other polarizable material, on the 
hydrophilic side, and in contact with vacuum (``air'') on the hydrophobic 
side. The work of Alper {\em et al}\cite{alper} focusses on the influence
of an amphiphilic layer on a water substrate, therefore the water is treated
explicitly. A few recent studies have also probed amphiphiles 
at oil/water interfaces\cite{marquez,urbina,schweighofer} with explicit 
oil molecules; much work clearly remains to be done here.
Rice and coworkers have investigated special systems, such as 
monolayers of fluorinated amphiphiles\cite{ricef} and monolayers
on Cs\cite{ricecs} and ice\cite{ricei}. 

Bilayers have received even more attention.  In the early studies, water 
has been replaced by a continuous medium like in the monolayer 
simulations\cite{berendsen1,khalatur,xiang,taga}.
Today's bilayers are usually ``fully hydrated'', {\em i.e.}, water is 
included explictly. Simulations have been done at constant
volume\cite{berendsen2,brickmann,robinson1,wilson,zhou,rusling} 
and at constant pressure or fixed surface 
tension\cite{shinoda1,chiu,klein,feller1,berendsen3,edholm3}.
In the latter case, the size of the simulation box automatically adjusts 
itself as to optimize the area per molecule of the amphiphiles in the 
bilayer\cite{at}. If the pressure tensor is chosen isotropic, bilayers 
with zero surface tension are obtained. Constant (positive) surface tension 
can be enforced by choosing an anisotropic pressure tensor, which is 
smaller in the directions parallel to the 
bilayer\cite{chiu,feller1,berendsen3}. A comparison of different
simulation methods, different ensembles, different water 
models, and different hydrocarbon models, has been made recently by 
Tielemann and Berendsen\cite{berendsen3}.

Whereas the main challenge for the first bilayer simulations has been to 
obtain stable bilayers with properties ({\em e.g.}, densities) which compare 
well with experiments, more and more complex problems can be tackled nowadays.
For example, lipid bilayers were set up and compared in different phases 
(the fluid, the gel, the ripple phase)\cite{klein,taga,berendsen2,shinoda2}.
The formation of large pores and the structure of water in these water
channels has been studied\cite{berendsen4,shinoda2}, and the forces
acting on lipids which are pulled out of a membrane have been
measured\cite{marrink}. 
The bilayer systems themselves are also becoming more complex. 
Bilayers made of complicated amphiphiles such as unsaturated lipids have been 
considered\cite{hyvonen,feller2}. The effect of adding cholesterol has been 
investigated\cite{edholm2,robinson2}. An increasing number of studies are 
concerned with the important complex of lipid/protein 
interactions\cite{edholm1,damodaran,shen} and, in particular, with the
structure of ion channels\cite{sansom,woolf,berendsen5}.

The third class of systems which have been investigated relatively extensively
by simulations in atomic detail are micelles. Again, the first studies
have ignored the water molecules and simply considered the conformations of 
amphiphiles which are confined into a shell of given spherical 
geometry\cite{haile,karaborni2}. Later studies focus on the stability of 
(preassembled) micelles in water, on the structure and mobility 
of the amphiphiles, and on the degree of water penetration into the 
hydrophilic shell and the hydrophobic 
core\cite{rusling,jonsson,watanabe1,wendoloski,laaksonen,mackerell}.
Reverse micelles have also been considered, as well in vacuum as
in an explicit oil environment\cite{griffiths,tobias}. Simulations
of more complex self-assembled structures are still in their infancy.
Watanabe and Klein have investigated cylindrical hexagonal phases using
simulation cells with appropriate boundary conditions\cite{watanabe2}. The 
structure of the cylinders turns out to be comparable to that of spherical
micelles in many respect (water penetration, conformational properties etc.).
More recently, Kong {\em et al}\cite{kong} have employed a simplified 
amphiphile model (ethoxy head group chains attached to a hydrocarbon 
continuum) to study the water-induced interactions between lamellae in a 
lamellar phase. They conclude that water restructuring plays an important 
part in stabilizing the lamellar phase.

\section{Idealized chain models}
\label{s3}

In coarse grained microscopic models, the amphiphiles, oil and water
molecules are still treated as individual particles, but their structure
is very much simplified. Molecular details are largely
lost. For reasons of computational efficiency, coarse grained
models are often formulated on a lattice. This obviously introduces
the danger of lattice artefacts; as we shall see, lattice models have
nevertheless been used successfully in the study of various phenomena and 
turn out to be extremely powerful. Off-lattice models are computationally 
more costly, but do not impose an {\em a priori} anisotropy on space, and 
are attracting growing interest. 

\subsection{Lattice models}
\label{s31}

The most complex and powerful coarse grained models are those which retain
the chain character of the amphiphile molecules.

In a class of ``realistic'' lattice models, hydrocarbon chains are placed 
on a diamond lattice in order to imitate the zigzag structure of the
carbon backbones and the trans and gauche bonds. Such models have been
used early on to study micelle structures\cite{pratt}, 
monolayers\cite{milik1}, and bilayers\cite{milik2}. Levine and coworkers have
introduced an even more sophisticated model, which allows to consider
unsaturated C=C bonds and stiffer molecules such as cholesterol: 
A monomer occupies several lattice sites on a cubic lattice, the saturated
bonds between monomers are taken from a given set of allowed bonds
with length $\sqrt{5}$, and torsional potentials are introduced to distinguish
between ``trans'' and ``gauche'' conformations\cite{levine1,levine2}. 

Most lattice models however abstract from the details of the hydrocarbon 
chain structure. The ``monomers'' of the model chains are then conceived
as effective monomers, which represent several (CH${}_2$) (or other) units 
in the molecules. One particularly popular lattice model has been introduced 
by Larson {\em et al}\cite{tanny,larson1}: In this model, oil and water 
molecules occupy single sites of a cubic lattice, and amphiphiles are chains 
made of ``tail'' monomers $T$ and ``head'' monomers $H$, which are identical 
to the oil and water particles, respectively. The amphiphile monomers are 
connected by bonds with one of the 26 nearest or diagonally 
nearest neighbors. Since the lattice is entirely filled with either oil, 
water, or amphiphile, one only has one independent (dimensionless) interaction
parameter $w/k_B T$, the relative repulsion between oil and water particles or
tail and head monomers. The interaction range is chosen such that particles
interact with their nearest neighbors and diagonally nearest neighbors,
{\em i.e.}, the coordination number is again 26.
The phase behavior of the model is solely controlled by the architecture of 
the amphiphile. It has been studied for a number of systems by Larson himself 
and others\cite{larson1,larson2,larson3,larson4,mackie} and turns out
to be amazingly multifarious. Perhaps the most spectacular achievement of
the model is that it seems to exhibit a gyroid 
phase\cite{larson4} (cf. Fig. \ref{fig2} (d)).
In a narrow concentration region, and if the size of the simulation
box matches closely the preferred spacing of the gyroid unit cell, the
amphiphiles self-assemble sponataneously into a gyroid upon cooling. This may 
not be conclusive evidence that the gyroid is actually the phase with the
lowest free energy in an infinitely extended system, but it nevertheless
demonstrates impressively the power of simple lattice models. 
Some phase diagrams\cite{larson4} for binary water/amphiphile systems are 
shown in Fig. \ref{fig5}. Their topology resembles that of experimental phase
diagrams, {\em i.e.}, the gyroid phase intrudes between the hexagonal 
phase and the lamellar phase, as an intermediate between a state where
the amphiphilic sheets have high local curvature (the hexagonal phase),
and one without local curvature (the lamellar phase).

The Larson model and Larson-type models have been widely used to study 
micelles\cite{larson3,stauffer1,bernardes1,brindle,care,maiti3,chris1,chris2,
nelson}, amphiphiles at oil/water interfaces\cite{chowdhury1,maiti1,maiti3}, 
bilayers\cite{care,bernardes2} and various other 
problems\cite{bernardes3,bernardes4,bernardes5,jin,maiti2}. 
The models differ from each other in the range of the interactions 
and in the treatment of the amphiphile monomers. 
Other than in Larson's original model, most authors include only nearest 
neighbor interactions, sometimes in combination with a different underlying
lattice\cite{nelson}. Some models stay with Larson's simplifying assumption 
that the amphiphile monomers are identical to oil or water 
particles\cite{chris1,chris2,nelson}, but most of them equip the amphiphiles 
with new types of particles, often even allowing for more than two different 
monomer species. A popular description
going back to Stauffer and coworkers\cite{stauffer1} and to
Bernardes and coworkers\cite{bernardes1,bernardes2} adopts the language
of the Ising model and assigns a spin $S=-1$ to oil particles,
a spin $S=+1$ to water particles, and an integer spin ranging between
-1 and +2 to surfactant monomers. The interaction energy is described 
by a Hamiltonian \mbox{${\cal H} = - \epsilon \sum_{<ij>} S_i S_j$}, where
the sum $<ij>$ runs over nearest neighbor sites on the lattice.
Bernardes has demonstrated, that amphiphile with the architecture 
$(+1,0,-1,-1,-1,-1)$ or $(-1,-1,-1,-1,0,1,1,0,-1,-1,-1,-1)$ 
(two-tailed amphiphiles) can spontaneously self-assemble into 
vesicles\cite{bernardes4,bernardes5}. A snapshot of such a
vesicle is shown in Fig. \ref{fig6}.

A particularly simple lattice model has been utilized by Harris and 
Rice\cite{harris2} and subsequently by Stettin {\em et al}\cite{stettin} 
to simulate Langmuir monolayers at the air/water interface:
Chains on a cubic lattice which are confined to a plane at one end.
Haas {\em et al} have used the bond-fluctuation model, a more sophisticated 
chain model which is common in polymer simulations, to study the same 
system\cite{haas1}. Amphiphiles are modeled as short chains of monomers,
which occupy a cube of eight sites on a cubic lattice, and are connected by
bonds of variable length\cite{bfm}. At high surface coverage, 
Haas {\em et al} report various lattice artefacts. They conclude that the 
study of dense surfactant monolayers calls for the use of off-lattice models, 
especially if one is interested in phenomena such as collective tilt etc. 
Nevertheless, the bond-fluctuation model can be a useful tool in other 
contexts. For example, M\"uller and Schick\cite{marcus} have employed it to 
study the formation and the structure of pores in amphiphilic bilayers.
From the bilayer undulations, they were able to extract the bilayer tension, 
and from the distribution of pore sizes, they could deduce the line tension 
of the pore edges. Fig. \ref{fig7} shows an example of a pore. It is
found to be ``hydrophilic'', {\em i.e.}, the amphiphiles around the
pore rearrange themselves so as to shield the bilayer core from the 
unfavorable solvent.

Last in this section on lattice chain models, let us cite the somewhat 
different approach of Jennings {\em et al}\cite{jennings}, who model the 
amphiphiles as single site particles on a lattice, but surround them with 
long hydrophobic chains (of chain length up to $N=80$). Their study 
focusses on the influence of amphiphiles on the conformations of nonpolar 
polymers. They report phenomena such as amphiphile induced polymer collapse 
and the stabilization of lamellar phases.

Lattice models have the advantage that a number of very clever Monte Carlo 
moves have been developed for lattice polymers, which do not
always carry over to continuum models very easily. For example,
Nelson {\em et al} use an algorithm which attempts to move vacancies 
rather than monomers\cite{nelson}, and thus allows to simulate the dense 
cores of micelles very efficiently. This concept cannot be applied to 
off-lattice models in a straightforward way. On the other hand, a number
of problems cannot be treated adequately on a lattice, especially 
those related to molecular orientations and nematic order. For
this reason, chain models in continuous space are attracting growing
interest.

\subsection{Chain models in continuous space}
\label{s32}

The usual structure of off-lattice chain models is reminiscent of
the Larson models: The water and oil particles are represented by spheres
(beads), and the amphiphiles by chains of spheres which are joined
together by harmonic springs 
\begin{equation}
 U_{ij}^b = k_b/2 \; (|\vec{r}_i-\vec{r}_j| - b)^2.
\end{equation}
Here a hard or smooth cutoff $b^c$ is sometimes imposed, such that
\begin{displaymath}
U_{ij}^b = \infty  \qquad \mbox{for} \qquad
b - b^c < |\vec{r}_i - \vec{r}_j| < b + b^c,
\end{displaymath}
in order to ensure that the beads cannot move arbitrarily far apart from 
each other. Spheres of type $i$ and $j$ often interact {\em via} truncated 
and shifted Lennard-Jones potentials
\begin{equation}
 U_{ij}^{LJ}(r) = \left\{ \begin{array}{rl}
4 \epsilon_{ij} 
\Big[ (\sigma_{ij}/r)^{12}-(\sigma_{ij}/r)^{6}) \Big] + C_{ij} 
& \qquad r < R_{ij}^c \\
0 & \qquad \mbox{otherwise}
\end{array} 
\right.,
\end{equation}
where $C_{ij}$ is chosen such that $U_{ij}^{LJ}(r)$ is continuous everywhere. 
In most cases the interaction parameter $\epsilon_{ij}$ is chosen
species-independent, $\epsilon_{ij}\equiv \epsilon$, and the
``sign'' of the interaction is controlled by the cutoff length
$R_{ij}^c$: Purely repulsive interactions between hydrophilic and hydrophobic 
units are obtained by choosing $R_{ij}^c=2^{1/6} \sigma_{ij}$.
For the attractive interactions, $R_{ij}^c = 2.5 \sigma$ (most common) or 
$ 2 \sigma$ is often used. 
In addition, a bending potential is sometimes introduced which
favors chain stretching and makes the chains stiffer. A model of
this kind has first been introduced by Smit {\em et al}\cite{smit1}
and later employed to explore the interplay of micelle formation and
amphiphile adsorption at an oil/water interface\cite{smit2,karaborni3}, 
the self-assembly of micelles in general\cite{smit3,smit4,palmer,goetz},
and that of bilayers\cite{goetz}. In the case of binary systems, the model
can be further simplified by ignoring the solvent particles. Even then,
one can still observe and study self-assembling micelles 
very well\cite{gottberg,bhattacharya,viduna}. 

The different surfactant models vary mostly in details.
Karaborni {\em et al}\cite{smit4} explore the influence of different
chain architectures on the micelle shapes. Inserting spacers between
hydrophobic tails in a double tailed amphiphile turns out to transform
former spherical micelles into threadlike micelles. (A similar result
was observed by Maiti and Chowdhury in the Larson model\cite{maiti3}.)
G\"otz and Lipowsky\cite{goetz} replace the repulsive interactions between
hydrophilic and hydrophobic beads by the softer potential 
$U = 4 \epsilon (\sigma/r)^9$ and introduce spontaneous tilt angles
in the bending potential in some of their simulations. Harries 
{\em et al}\cite{benshaul2} and von Gottberg {\em et al}\cite{gottberg} 
connect the beads by rigid rods, {\em i.e.}, choose an infinitely large 
spring constant $k_b$. 
Bhatthacharya {\em et al}\cite{bhattacharya} compare nonionic surfactants with 
ionic surfactants and model the latter by including an additional Yukawa-type 
electrostatic interaction $U(r) \propto \exp(-\kappa/r)$ between the head 
groups. Perhaps not surprisingly, the charged micelles show a much stronger
ordering tendency than the neutral micelles. 

A few groups replace the Lennard-Jones interactions by interaction of
a different form, mostly ones with a much shorter interaction 
range\cite{viduna,baumgartner1,baumgartner2}. 
Since most of the computation time in an off-lattice simulation is
usually spent on the evaluation of interaction energies, such a measure
can speed up the algorithm considerably. 
For example, Viduna {\em et al} use a potential, in which the interaction 
range can be tuned,
\begin{equation}
 U^M(r) = E_{ij} \Big[ e^{-2 \alpha (r-\sigma)}-2 e^{-\alpha(r-\sigma)} \Big]
 - C \qquad \mbox{for $r < R_c$}
\end{equation}
(Morse potential). The constant $C$ is defined such that $U^M(r)=0$ at the 
cutoff distance $R_c$. The interaction range is determined by the parameter 
$\alpha$, which Viduna {\em et al} choose very large, $\alpha=24$.
Hence the cutoff distance can be made small 
($R_c = 1.25 \sigma$ in \cite{viduna}). This model has first been used
by Gerroff {\em et al}\cite{gerroff} and is discussed in some detail 
in chapter 9 of this book.

Bead spring models without explicit solvent have also been used
to simulate bilayers\cite{benshaul2,baumgartner1,baumgartner2} and
Langmuir monolayers\cite{haas2,haas3,christoph1,christoph2,opps}. 
The amphiphiles are then forced into sheets by tethering the head
groups to two dimensional surfaces, either {\em via}
a harmonic potential, or {\em} via a rigid constraint.

Baumg\"artner and coworkers\cite{baumgartner1,baumgartner2} study
lipid-protein interactions in lipid bilayers. The lipids are modeled
as chains of hard spheres with heads tethered to two virtual surfaces, 
representing the two sides of the bilayer. Within this model,
Baumg\"artner\cite{baumgartner1} has investigated the influence
of membrane curvature on the conformations of a long embedded chain 
(a ``protein''). He predicts that the protein spontaneously
localizes on the inner side of the membrane, due to the larger fluctuations
of lipid density there. Sintes and Baumg\"artner\cite{baumgartner2} have
calculated the lipid-mediated interactions between cylindrical inclusions 
(``proteins''). Apart from the usual depletion interaction at contact,
they find that the lipids induce a net attractive force between the
inclusions over the much wider range of $\sigma <r< 6 \sigma$, where
$\sigma$ is the diameter of a lipid bead.

Simulations of monolayers have focussed on internal phase transitions, 
{\em e.g.}, between the expanded phase and the condensed phases, 
between different tilted phases etc. These phenomena cannot be reproduced
by models with purely repulsive interactions. 
Therefore, Haas {\em et al}\cite{haas2,haas3} represent the amphiphiles as 
stiff Lennard-Jones chains, with one end (the head bead) confined to move 
in a plane. In later versions of the model\cite{christoph1,christoph2,opps}, 
the head bead interactions differ from those of the tail beads: They are 
taken to be purely repulsive, and the head size is variable. 

Stadler {\em et al} have performed Monte Carlo simulations at constant 
pressure of this model, and calculated the phase behavior for various 
different head sizes. It turns out to be amazingly rich. 
The phase diagram for chain length $N=7$ and heads of size $1.2 \sigma$ 
($\sigma$ being the diameter of the tail beads) is shown in Fig. \ref{fig8}.
A disordered expanded phase is found as well as several condensed phases
with different tilt order -- a phase without collective tilt, one with tilt 
towards nearest neighbors, and one with tilt towards next nearest neighbors.
Particularly unexpected is a phase with a superstructure (LC-mod), where the 
direction of tilt is modulated and points on average towards an intermediate
between nearest neighbors and next nearest neighbors. The heads in the
condensed phases are arranged on a hexagonal lattice, which is distorted
in the direction of tilt. In order to avoid shear stress in the system,
it is thus crucial to let fluctuate not only the size of the simulation box, 
but also its shape. Moreover, huge hysteresis effects were observed
at some of the phase boundaries, and the free energies of the competing 
phases had to be calculated by thermodynamic integration methods\cite{como} 
in order to locate the phase boundaries. Configuration snapshots of the 
expanded phase and the modulated condensed phase are shown in Fig. \ref{fig9}.

The modulated phase disappears for head sizes smaller than $1.14 \sigma$. 
Modulations of this kind have not been observed in experiments so far. 
However, Stadler {\em et al}
argue that the intensity of the satellite peaks which would be indicative
of the superstructure is so low, that they could not possibly be detected 
in X-ray measurements with the usual experimental resolution. Phases with 
intermediate tilt directions have been reported\cite{mono_reviews,kaganer}.

To complete this overview over chain models, we mention the dimer models,
which represent the amphiphiles by just two units attached to each
other\cite{baumgartner3,rector,laradji1,kuespert,bolhuis}. They have been
used to study curved bilayers\cite{baumgartner3}, the kinetics 
of phase separation between oil and water at the presence of
surfactants\cite{laradji1}, and some aspects about self-assembled 
micelles\cite{rector,bolhuis} (see below).

\subsection{An application: Micelle shapes and size distributions}
\label{s33}

As has probably become obvious already, the study of micelles has been
one of the big topics in simulations of systems with surfactants. 
We have cited many of the related publications in the previous
sections. Here, we shall discuss some special aspects of micelle
simulations in order to illustrate the use of idealized chain models
for this type of problem. 

The first important step in a micelle simulation obviously consists in 
defining a ``micelle''. In simulations of systems with short range 
interactions, this can be done in a relatively straightforward way: 
Molecules are said to form an aggregate if they are in contact with each 
other, {\em i.e.}, if they interact. After having set this straight, one 
can proceed to measure the concentration of {\em free}, {\em i.e.}, 
noninteracting surfactants as a function of the total surfactant 
concentration. In self-assembling systems, it saturates very soon after 
a first regime of linear increase, and stays constant or even decreases 
slightly\cite{gottberg} thereafter. Similarly, the chemical potential of 
the surfactant as a function of the surfactant concentration exhibits a 
marked kink, and turns from initial increase almost into a 
constant\cite{rector,chris2}. The concentration at which this
happens is called the critical micelle concentration (CMC). 
The sharp changes at the CMC can be understood from the competition between 
the translational entropy of the free surfactants and the energy, which their
hydrophobic tails gain in the micelles due to the reduced number of 
unfavorable water contacts. Let us consider surfactants which are part of an 
aggregate with the aggregation number $N$, and neglect the interactions
between aggregates. Since the aggregates are at equilibrium with each other, 
the chemical potential at temperature $T$ can be written as\cite{israelachvili}
\begin{equation}
\label{mu}
\mu = \mu_N^0 + k_B T/N \ln(f_N X_N/N)
\end{equation}
for all $N$, where $\mu_N^0$ denotes the mean interaction free energy in the 
aggregate, and the second term describes its translational entropy.
Here $X_N$ is the volume fraction of amphiphiles in an aggregate of size $N$,
and the activity coefficient $f_N$ is of order one. 
From eqn. (\ref{mu}) one concludes 
immediately
\begin{equation}
\label{xn}
f_N X_N = N  \Big[ f_1 X_1 e^{(\mu_1^0-\mu_N^0)/k_B T} \Big]^N.
\end{equation}
Note that $\mu_1^0 > \mu_N^0$, since the surfactants gain energy in the 
aggregates. Large aggregates can therefore be expected to dominate at 
surfactant concentrations larger than
\begin{equation}
(X_1)_{CMC} \approx e^{(\mu_N^0-\mu_1^0)/k_B T}.
\end{equation}
Beyond the CMC, surfactants which are added to the solution thus form 
micelles, which are at equilibrium with the free surfactants. This explains
why $X_1$ and $\mu$ level off at that concentration. Note that even though
it is called ``critical'', the CMC is not related to a phase transition. 
Therefore, it is not defined unambiguously. In the simulations, some authors 
identify it with the concentration where more than half of the surfactants 
are assembled into aggregates\cite{stauffer1}; others determine the
intersection point of linear fits to the low concentration and the high 
concentration regime, either plotting the free surfactant concentration vs. 
the total surfactant concentration $\Phi$\cite{bernardes1}, or
plotting the surfactant chemical potential vs. $\ln (\Phi)$\cite{chris2}.

We turn to the discussion of micelle size distributions and micelle
shapes. In order to proceed with the analysis, one needs an expression for 
the interaction free energy $\mu_N^0$ of surfactants in aggregates of
size $N$. Following Israelachvili\cite{israelachvili}, we assume
that the amphiphiles pack closely into aggregates, occupying a fixed volume
per amphiphile $v$, and that $\mu_N^0$ is simply given by the surface
free energy of the aggregates:
\begin{equation}
\label{mn}
\mu_N^0 = \frac{\gamma}{a} \; (a-a_0)^2 + \bar{\mu}^0,
\end{equation}
where $\gamma$ is the interfacial tension, $a$ the surface area per
molecule, and $a_0$ the optimal head group area. In the case of spherical
micelles, the surface area is given by $a = (36 \pi v^2)^{1/3} N^{-1/3}$.
The free energy (\ref{mn}) has a minimum at the aggregation number 
$M=36 \pi v^2/a_0^3$, and can be expanded around this minimum to yield
\begin{equation}
\mu_N^0 = \mbox{const.} + \frac{\gamma a_0}{9 M^2} \; (N-M)^2. 
\end{equation}
Inserting this into eqn. (\ref{xn}), one obtains a size distribution
for spherical micelles which is approximately Gaussian, 
\begin{equation}
\label{xns}
X_N^S \propto  e^{-(N-M)^2/2 \sigma^2} \qquad \mbox{with} \qquad
\sigma = \sqrt{\frac{9 k_B T M}{2 \gamma a_0}}.
\end{equation}
Hence the sizes of spherical micelles are distributed around a most
probable aggregation number $M$, which depends only on molecular details
of the surfactants in this simplest approximation. Indeed, micelle size
distributions at concentrations beyond the CMC have shown a marked peak 
at a given aggregation number in many simulations\cite{chris1,larson3,
stauffer1,care,chris2,smit3,viduna,rector,bolhuis}.

Larger aggregates have seldom spherical geometry, but tend to form
cylindrical micelles. In this case, the diameter of the cylinders can
usually be adjusted such that the head groups can cover their optimal
head group area $a_0$, and the interaction free energy per surfactant
reduces to the constant $\bar{\mu}^0$. The size distribution for
cylindrical micelles is then exponential in the limit of large $N$,
\begin{equation}
\label{xnc}
X_N^C \propto  e^{-N \alpha} \qquad \mbox{with} \qquad
\alpha = - \ln(f_1 X_1) - (\mu_1^0 - \bar{\mu}^0)/k_B T.
\end{equation}
At small $N$, correction terms come into play which account for the ends
of the cylinders. In particular, the aggregation number of cylindrical 
micelles in this simple picture must always be larger than $M$, the most
probable aggregation number of a spherical micelle. Putting everything
together, the expected size distribution has a peak at 
$M$ which corresponds to spherical micelles, and an exponential 
tail at large $N$ which is due to the contribution of cylindrical micelles. 

That this is indeed so, has been demonstrated nicely by Nelson 
{\em et al}\cite{nelson} in extensive simulations of the Larson model. 
Fig. \ref{fig10} shows an example of a micellar size distribution for $H_2T_2$ 
surfactants in an aqueous environment, at 7.5 \% surfactant volume fraction. 
The main peak is fitted to a Gaussian distribution (\ref{xns}), and the tail 
to an exponential distribution of the form (\ref{xnc}),
$X_N \propto f(N) \exp(-\alpha N)$, where the prefactor $f(N)$ interpolates 
smoothly between $f(N)=0$ at $N=M$ and $f(N)=1$ at $N \gg M$. 
The results confirm clearly the predicted exponential decay at large $N$.
The shape of the micelles can be studied directly by comparing the 
different eigenvalues of the radius of gyration tensor of the aggregates.
Nelson {\em et al} find that they are approximately equal, as long as
the aggregation number is smaller than the most probable size 
$M \approx 40$. Beyond that value, the largest eigenvalue grows linearly
to become much larger than the other two. This observation is consistent
with the picture that the peak of the distribution is generated by
spherical micelles, and the tail by spherically capped cylindrical micelles.

Whereas the evolution from spherical to cylindrical shapes as a function
of aggregation number appears to be gradual in the simulations of Nelson 
{\em et al}, Viduna {\em et al} report a rather dramatic transition at a 
well-defined aggregation number in their off-lattice simulations of 
$H_2 T_2$ micelles. The eigenvalues of the radius of gyration tensor are 
shown as a function of the aggregation number in Fig. \ref{fig11}. At this 
choice of parameters, the micelle size distribution exhibits a peak at 
$N \approx 28$. The shape parameters seem rather unaffected by this, 
one barely notices that the largest eigenvalue begins to increase a little 
bit faster than the other two. In contrast, at $N \approx 47$, it 
rises abruptly and attains a larger value, where it seems to level
off again. This interesting ``shape transition'' clearly deserves
further investigation in the future.

Other questions which will have to be studied in more detail refer
to the role of micelle interactions. Note that they are not included
in the approximate analysis sketched above. Von Gottberg 
{\em et al}\cite{gottberg} have argued that excluded volume interactions 
between micelles are responsible for the slight decrease of $X_1$ at higher
surfactant concentrations, which has been observed in many 
simulations\cite{chris2,gottberg,viduna}. Bolhuis and Frenkel\cite{bolhuis}
have studied the influence of intermicellar interactions on the micelle
size distributions within a simple model, which approximates
the micelles as hard spheres of variable size. Their results indicate
that the interactions affect the size distribution in a way which is
equivalent to shifting the chemical potential to an effectively lower
value.

\section{Lattice spin models and others}
\label{s4}

Whereas chain models still allow for a relatively unified treatment
of various aspects about amphiphilic systems, such as their bulk phase 
behavior and the properties of monolayers and bilayers, this is not
true any more for the even more idealized models at the next level
of coarse graining. These usually have to be adapted very specifically
to the problem one wishes to study. 

One particularly favored class of models have been models of Ising type,
which represent the particles by states on sites or bonds of a lattice. 
Those intended to describe bulk amphiphilic systems have been reviewed 
recently by Gompper and Schick in Ref. \cite{gs_review}, and those developed 
for monolayers and bilayers by Dammann {\em et al} in Ref. \cite{bm_review}.  
Hence we shall only briefly recall their main features here, and focus
on the discussion of their use in computer simulations.

\subsection{Bulk systems}
\label{s41}

Lattice models for bulk mixtures have mostly been designed to describe 
features which are characteristic of systems with low amphiphile content. 
In particular, models for ternary oil/water/amphiphile systems are challenged 
to reproduce the reduction of the interfacial tension between water and oil 
at the presence of amphiphiles, and the existence of a structured disordered 
phase (a microemulsion), which coexists with an oil rich and a water rich 
phase. We recall that a ``structured'' phase is one in which correlation
functions show oscillating behavior. Ordered ``lamellar'' phases have also 
been studied, but they are much more influenced by lattice artefacts 
here than in the case of the chain models.

The most senior among the lattice models is the Widom model, which has 
been formulated already in 1968 in its first version by Wheeler and 
Widom\cite{widom}.
The water, oil and amphiphile molecules are represented by (1,1), (-1,-1) 
and (1,-1) {\em bonds}, respectively, on a two state Ising lattice. 
Hence oil and water are always separated by amphiphiles by construction. 
If the particles are taken to be non-interacting otherwise, the model maps 
directly onto the Ising model with nearest neighbor interactions, where the 
role of the temperature is assumed by the chemical potential of the 
amphiphile. The resulting phase diagram is well-known, one finds
a region of oil/water coexistence and a critical point. In order to obtain 
more complex phase behavior, interactions between particles have to be 
added. For example, a penalty is often imposed on amphiphiles which meet at 
their one end. This introduces additional terms into the equivalent Ising 
Hamiltonian, 
\begin{equation}
\label{widom}
{\cal H} = - h \sum_i \sigma_i
- J \sum_{ij} \sigma_i \sigma_j 
- 2M \sum'_{ij} \sigma_i \sigma_j
- M \sum''_{ij} \sigma_i \sigma_j 
\end{equation}
where the second sum runs over nearest neighbors, the third over next
nearest neighbors, and the fourth over fourth nearest neighbors. The first 
two terms drive the volume fraction of oil, water and amphiphile in the 
system, the third term incorporates some sort of ``bending energy'' of the 
amphiphilic sheets, and the last term an interaction between sheets.
The model exhibits a region of coexistence between oil and water rich
phases, a region with various ordered phases, and a disordered structured
phase. In the simple version (\ref{widom}), the transition between the 
oil/water region and the disordered region is continuous, hence the model 
does not recover three phase coexistence. However, this can be remedied by 
including suitable additional bond interactions\cite{hansen}.
The phase behavior of the model has been investigated in detail in 
Monte Carlo simulations, in particular by Stauffer and 
coworkers\cite{hansen,stauffer2,dawson}. Chowdhury and coworkers
have used it to study various other aspects of amphiphilic systems, 
such as the effect of confinement on a microemulsion\cite{chowdhury2},
the lifetime, stability and rupture of Newton black films\cite{chowdhury3},
and the roughness of an amphiphilic film\cite{maiti4}.

Even though the basic idea of the Widom model is certainly very appealing,
the fact that it ignores the possibility, that oil/water interfaces are 
not saturated with amphiphiles, is a disadvantage in some respect. 
The influence of the amphiphiles on interfacial properties cannot be studied 
on principle; in particular, the reduction of the interfacial tension cannot 
be calculated. In a sense, the Widom model is not only the first microscopic 
lattice model, but also the first random interface model: Configurations are 
described entirely by the conformations of their amphiphilic sheets.

As an alternative, Alexander\cite{alexander} has proposed a model
which places the oil and the water on lattice sites and distributes
the amphiphiles on the bonds. The interactions between two neighboring
oil and/or water particles depend on whether or not the connecting bond is
occupied by an amphiphile. As long as the amphiphiles themselves do not 
interact with each other, the sum over possible bond states can be carried 
out independently, and the model turns out to be equivalent to an Ising
model with temperature dependent interactions. Hence this simplest version
does not capture the specific properties of amphiphilic systems, and 
additional amphiphile interactions have to be included. A variety of 
different interaction terms have been implemented: bending energies, edge 
energies, corner energies, special penalties if amphiphilic sheets meet or 
cross each other etc. Different versions of the model have been explored in 
Monte Carlo simulations by Ebner and coworkers\cite{ebner} and
Stockfisch and Wheeler\cite{stockfisch}. The model displays lamellar
phases, and three phase coexistence between an oil rich, a water rich
and a disordered phase with intrinsic structure. 
Moreover, the amphiphiles were shown to reduce the interfacial tension
between oil and water by a factor of up to 1000\cite{stockfisch}.
The interfacial tension was determined with the histogram method of 
Binder\cite{binderh}, which shall be discussed in more detail below.

More recently suggested models for bulk systems treat oil, water and
amphiphiles on equal footing and place them all on lattice sites.
They are thus basically lattice models for ternary fluids, which are
generalized to capture the essential properties of the amphiphiles.
Oil, water and amphiphiles are represented by Ising spins $S = -1, 0$ and 
+1. If one considers all possible nearest neighbor interactions between these
three types of particles, one obtains a total number of three independent
interaction parameters, and two independent chemical potentials. In Ising
language, the most general Hamiltonian then reads
\begin{equation}
{\cal H} = - \sum_{ij} \Big[
J S_i S_j + K S_i{}^2 S_j{}^2 + C(S_i{}^2 S_j + S_j{}^2 S_i ) \Big]
- \sum_i \Big[ H S_i - \Delta S_i{}^2 \Big],
\end{equation}
which is exactly the Hamiltonian of the Blume-Emery-Griffiths 
model\cite{beg}. Balanced systems with intrinsically identical oil and water 
particles are described by the set of parameters $C=0$ and $H=0$. 

The first extension of this model, which accounts for the special character 
of the amphiphiles, has been the three component model, introduced by Schick 
and Shih in 1987\cite{3comp}. They simply add an additional triplet 
interaction
\begin{equation}
{\cal H}_{amp} = - L \sum_{ijk} S_i (1-S_j^2) S_k,
\end{equation}
between triplets $(ijk)$ of sites in a line. At $L<0$, this term imposes a 
energy penalty if an amphiphile is sitting between two oil or two water 
particles, and offers an extra energy reward if an amphiphile separates oil 
from water. The model exhibits a structured disordered phase, which may 
coexist with an oil-rich and a water-rich phase in three dimensions, 
and ordered ``lamellar'' phases. It has been examined in detail by mean 
field theories and other analytical methods\cite{gs_review}, but relatively 
seldom by computer simulations\cite{gompper1,friederike4}. 

In order to illustrate the type of questions which can be addressed within
such an idealized lattice model, we show a histogram of the normalized order 
parameters $M=\sum_i S_i/N$ and $Q=\sum_i S_i{}^2/N$ ($N$ being the system 
size) in Fig. \ref{fig12}\cite{friederike4} for a point in phase space, 
where oil- and water-rich phases coexist with a structured microemulsion, 
The peaks corresponding to the three phases are readily identified at
$M \approx \pm 1$ and $M=0$. Moreover, one discerns a broad band at 
$Q \approx 0.94$ which extends over nearly the whole range of $M$. 
The story behind this plot is the following: Based on mean field arguments,
Gompper and Schick\cite{gompper2} had made the prediction that a structured
microemulsion should not wet the oil/water interface. This is indeed
observed in systems with strongly structured microemulsions; however, 
it is not always true for weakly structured microemulsions\cite{strey4}. 
One reason had already been pointed out by Gompper and Schick\cite{gompper2}: 
long range van der Waals forces shift the wetting transition beyond the 
structured side of the disorder line. Other factors which usually influence 
phase transitions are fluctuations. The Monte Carlo study\cite{friederike4} 
aimed to elucidate this latter aspect. 

Having obtained a set of histograms like Fig. \ref{fig12} for different system
sizes, (using histogram reweighting methods\cite{histo} in larger systems),
the procedure was as follows: Assuming that the main contribution to the 
valleys between the peaks arises from configurations which contain two 
interfaces separating one phase regions, the interfacial tension between 
those phases is given 
by\cite{binderh}
\begin{equation}
 \gamma/k_B T = - \frac{1}{2A} \; \ln(N_{min}/N_{max}) 
\end{equation}
where $A$ is the interfacial area. The interfacial tensions $\gamma_{om}$
and $\gamma_{wm}$ between the oil or water rich phase and the microemulsion 
can thus be extracted directly from the valleys between the peaks at 
$M=\pm 1$ and $M=0$. Furthermore, one estimates the number of amphiphiles 
needed to form just two sheets of amphiphiles, and realizes that the 
corresponding configurations are just those which belong to the band at 
$Q=0.94$. 
Hence the direct interfacial tension $\gamma_{ow}$ between the oil- and the 
water-rich phase can be calculated from the area under that band. 
The microemulsion wets the oil/water interface if 
$\gamma_{ow} < \gamma_{om} + \gamma_{wm}$. Note that the numbers obtained
for the interfacial tensions with this method are subject to strong finite 
size corrections, which have to be analyzed carefully.

For the system studied in \cite{friederike4}, it turns out that the oil/water 
interface is not wetted by the microemulsion, even though the latter is 
weakly structured. Hence fluctuations do shift the wetting transition beyond 
the disorder line. This has been explained later by Schmid and 
Schick as the effect of capillary wave fluctuations of the interface
positions\cite{friederike5}. The amphiphiles in \cite{friederike4} are found 
to reduce the total interfacial tension by a factor of 100, a value which is 
consistent with experimental results for comparably weak 
amphiphiles\cite{strey4}.

The example illustrates, how Monte Carlo studies of lattice models can
deal with questions, which reach far beyond the sheer calculation of
phase diagrams. The reason why our particular problem could be studied with 
such success lies of course in the fact that it touches a rather fundamental 
aspect of the physics of amphiphilic systems -- the interplay between 
structure and wetting behavior. In fact, the results should be universal and 
apply to all systems where structured, disordered phases coexist with 
non-structured phases. It is this universal character of many issues in 
surfactant physics, which makes these systems so attractive for 
theoretical physicists.

After this short intermezzo, we turn back to introduce the last class
of lattice models for amphiphiles, the vector models. Like the three
component model, they are based on the three state Ising model for ternary
fluids; however, they extend it in a way that they account for the 
orientations of the amphiphiles explicitly: Amphiphiles (sites with $S=0$) 
are given an additional degree of freedom $\vec{\tau}_i$, a vector with 
length unity,
which is sometimes constrained to point in one of the nearest neighbor
directions, and sometimes completely free. It is set to zero on sites
which are not occupied by amphiphiles. A possible interaction term which
accounts for the peculiarity of the amphiphiles reads
\begin{equation}
\label{hav}
{\cal H}_{amph} = - J \sum_{ij} 
(\sigma_i \vec{\tau}_i \vec{r}_{ji} + \sigma_j \vec{\tau}_i \vec{r}_{ij}),
\end{equation}
where $\vec{r}_{ij}$ denotes the vector which connects the site $i$ to
the site $j$. The new interaction thus awards an energy bonus if an 
amphiphile points towards a water molecule or away from an oil molecule. 
As long as the $\tau_i$ do not interact with each other, one can
get rid of them by integrating them out exactly. However, this generates
a set of temperature dependent multiplet interactions, which are not 
necessarily easier to handle\cite{gs_review,slotte}. Probably the first
Monte Carlo simulation of such a model (for a binary water/amphiphile system)
has been performed by Halley and Kolan\cite{halley} in 1988. Later, ternary 
systems were studied in two dimensions by Slotte\cite{slotte} and
Laradji {\em et al}\cite{laradji2}, and in three dimensions by
Gunn and Dawson\cite{gunn1}, Matsen {\em et al}\cite{mark1} and
Linhananta\cite{linhananta}. Fig. \ref{fig13} shows a phase diagram
of Matsen {\em et al}. As the previously discussed models,
vector models display lamellar phases, regions of oil/water coexistence
and a structured microemulsion. Fluctuations are found to have a particularly 
destabilizing effect on the lamellar phase\cite{laradji2,mark1}, and 
to increase, in turn, the region of stability of the microemulsion.
This has a significant effect on the topology of the phase diagram in
both two and three dimensions. In two dimensions, phase coexistence
between the lamellar phase and the oil- and water-rich phases is
entirely suppressed, and the lamellar phase is separated from
the oil/water coexistence region by a microemulsion channel at all
temperatures. Matsen\cite{mark2} has conjectured that this should generally
be the case for microemulsion models in two dimension, and argued that as a 
consequence, two dimensional models also fail to produce a triple line of 
coexistence between an oil-rich, a water-rich and a disordered phase, as soon
as the surfactant is efficient enough to produce a lamellar phase.
Hence fluctuations suppress three phase coexistence of oil, water and 
microemulsions in two dimensions. In three dimensions, the effect is just 
opposite.  The mean field phase diagram corresponding to Fig. \ref{fig13} does
not exhibit three phase coexistence. By stabilizing the microemulsion, the
fluctuation unveil a region of three phase coexistence, which would
otherwise be covered by the coexistence region between the lamellar phase
and the oil/water-phases.

Note that Fig. \ref{fig13} has still little similarity with the experimental
phase diagram of ternary amphiphilic systems, Fig. \ref{fig3}b. In 
particular, the region of three phase coexistence is not confined between
an upper and a lower critical endpoint, as in the experiments. This
is however not surprising, since the lower critical endpoint is probably
caused by orientational ordering in the water molecules, {\em i.e.},
the increasing structure of the hydrogen bond network. 
In order to account for this effect, Matsen {\em et al}\cite{mark3}
have proposed a vector model which attaches vector degrees of freedom
to both the amphiphile and the water molecules. They were able to
calculate a mean field phase diagram which is indeed remarkably similar
to the experimental phase diagram for weak amphiphiles. The model has
not been studied in Monte Carlo simulations so far.

A somewhat different type of vector model has been studied by Emerton,
Boghosian and coworkers\cite{coveney} in a set of recent papers. 
It is a lattice gas model, {\em i.e.}, the sites of a (triangular) 
lattice can also be empty. Amphiphiles are assigned a dipole vector,
and all particles are given a ``velocity'' $\vec{c}_i$, which points
in one of the lattice directions. The dynamics of the system is described
by collision processes: Particles move along the direction of their velocity,
collide, and are redistributed according to a Boltzmann weight under
consideration of various conservation laws. The interaction energies are
fairly complicated and shall not be spelled out here. The basic feature
of the model is that it includes hydrodynamic interactions and allows
to study dynamical phenomena under conditions of conserved momentum.
At equilibrium, it exhibits the usual phases, lamellae, droplets and
bicontinuous structures. It has mainly been used to investigate
nonequilibrium phenomena, such as the kinetics of phase separation,
shear induced phase transitions etc.

Before moving on to the bilayer and monolayer models, we briefly discuss
a few idealized microemulsion models in continuous space. In general, 
off-lattice models on this level of coarse graining have attracted much 
less interest than lattice models.
Gunn and Dawson\cite{gunn2} have studied a mixture of Lennard-Jones spheres 
(water) and Gay-Berne ellipsoids (amphiphile) as a model for a binary 
amphiphile/water mixture (The Gay-Berne potential is a distorted 
Lennard-Jones potential). An additional interaction similar to (\ref{hav})
between amphiphiles and water is added in order to mimic the amphiphilic
nature. The system is simulated under constant pressure and constant 
temperature conditions. It exhibits a crystalline lamellar phase, a
fluid lamellar phase, and a disordered liquid crystal. Unfortunately,
Gunn and Dawson do not seem to have pursued their studies of this model.
Drouffe {\em et al}\cite{drouffe} have studied self assembly into
two dimensional layers using a two dimensional model of hard spheres, 
which are decorated with interacting intrinsic ``orientations'' $\vec{n}$. 
Two dimensional aggregation into ring vesicles has also been examined by 
Saito and Morikawa\cite{saito} within a hard rod model. Recently, 
de Miguel and Telo da Gama\cite{miguel} have introduced a model which borrows
elements from Drouffe's model and from the lattice vector models:
Particles are represented by hard spheres of equal diameter $\sigma$,
with attractive (species dependent) van der Waals potentials.
In addition, amphiphiles are given an extra dipole moment, which interacts
with water and oil particles with a potential which is again reminiscent
of the vector model potentials (\ref{hav}),
\begin{equation}
V_{ani} = \pm \epsilon (\sigma/r)^6 (\vec{n} \cdot \vec{r}/r).
\end{equation}
The model has not been studied very intensely so far, in particular,
none of the features which are characteristic for amphiphilic systems
have been recovered yet. However, it is close enough to the successful
vector models and simple enough that it might be a promising candidate 
for off-lattice simulations of idealized amphiphilic systems in the future.

\subsection{Bilayer and monolayer models}
\label{s42}

Whereas microscopic models for bulk systems incorporate the amphiphilic
character and often the orientational properties of the surfactants
as basic ingredients, models for bilayers and monolayers are constructed 
to reproduce internal transitions, such as the gel-fluid transition, and 
therefore concentrate on rather different aspects of the surfactant structure.

For example, Scott {\em et al}\cite{scott} take interest in the ripple
phase in bilayers and have constructed a lattice model which assigns
two integer degrees of freedom to each site (``lipid'') on the lattice:
One Ising spin $\sigma = \pm 1$, which describes possible orientations
of the head group, and one integer $n \in \{0,\pm1,\pm2,\cdots\}$, which
represents the displacement of the molecules perpendicular to the
bilayer plane. 
Using a complicated Hamiltonian, borrowed from the chiral clock model, 
they indeed find a ripple phase with properties which are in decent agreement 
with experimental data. The interplay of head group orientation and ripple 
formation in lipid bilayers has also been studied by Schneider and Keller 
within a coarse-grained lipid model\cite{schneider}.

Other studies have been concerned with transitions between condensed
phases in Langmuir monolayers. Tilting phase transitions have been studied 
in some detail within models of grafted rigid rods\cite{chen,kreer,martin}.
Swanson {\em et al}\cite{swanson} study a fluid of up to 4096 particles 
with fourfold symmetry in order to model rotator phase transitions. These
are transitions from a state where the amphiphile tails are locked into 
each other, to a state where they are free to rotate around. Interestingly, 
a hexatic phase is found, in which positional correlations decay
exponentially, but the correlations between the orientations of the
bonds connecting nearest neighbors decay only algebraically. Many condensed 
monolayer phases are indeed believed to be hexatic\cite{mono_reviews}.

Most monolayer and bilayer studies focus on the gel-fluid transition, 
where the internal (conformational) degrees of freedom of lipids are 
important. In the most widely studied lattice model, the ten-state 
Pink model\cite{pink}, they are built in as intrinsic degeneracies. 
The Pink model assigns one of ten states, denoted $m$, to every site (lipid) 
on a triangular lattice. Every state corresponds to a set of conformations:
The lowest state $m=1$ to the (unique) all-trans configuration, the
states $m=2-9$ to almost ordered chains with a few isolated chain defects,
and the highest state $m=10$ to a completely disordered chain.
Every state is characterized by its degeneracy $D_m$ ($D_1=1$,
$D_m \approx 4-100$ for $m=2-9$,$D_{10} \approx 400.000$),
its energy $\epsilon_m$, and the area $A_m$ covered by the lipids. The
latter is approximated by $1/d_m$, the inverse chain length in the
state $m$. With this identification, $D_m$, $\epsilon_m$, and $A_m$ are 
single chain properties which can be calculated within an appropriate 
microscopic chain model. One further defines the occupation variable 
${\cal L}_{im}$, which takes the value 1 if the lipid $i$ is in the state $m$,
and 0 otherwise. The Pink Hamiltonian is then written as
\begin{equation}
{\cal H}_{pink} = 
\sum_i \sum_{m=1}^{10} (\epsilon_m + \Pi A_m) {\cal L}_{im} - \frac{J_0}{2} 
\sum_{ij} \sum_{m,n=0}^{10} I_m I_n {\cal L}_{im} {\cal L}_{jn},
\end{equation}
where $\Pi$ is an effective lateral pressure, $J_0$ the strength of the
van der Waals interaction between neighbor chains, and $I_m I_n$ an 
interaction matrix which accounts for the conformations of the chains
and the distance between two chains, and can also be calculated from
microscopic single chain properties. In practice, the model parameters have
often been chosen so as to describe phospholipid monolayers and bilayers.
The interactions between the monolayers in bilayer sheets are usually 
ignored. If one is not interested in a quantitative comparison with
experiments, one can choose to study a simplified version of the model.
For example, the two-state Doniach model\cite{doniach} distinguishes between 
only two states, one ordered state with $\epsilon_0=0$, and one highly
degenerate state with $\epsilon_1>0$. The interactions are such that
only ordered lipids are capable of interacting with each other.
A similarly simple model has been studied by Jerala {\em et al}\cite{jerala}.

The Pink model is found to exhibit a gel-fluid transition for lipids with 
sufficiently long chains, which is weakly first order. The transition
disappears in bilayers of shorter lipids, but it leaves a signature in
that one observes strong lateral density fluctuations in a narrow temperature
region\cite{mouritsen1,corvera}. In later studies, the model has been
extended in many ways in order to explore various aspects of gel-fluid
transitions\cite{lemmich}. 
For example, Mouritsen {\em et al}\cite{mouritsen2} have 
investigated the interplay between chain melting and chain crystallization
by coupling a two-state Doniach model or a ten-state Pink model to
a Potts model. (The use of Potts models as models for grain
boundary melting has been suggested by Sahni {\em et al} in 1983\cite{sahni}: 
Every Potts state is then identified with a different domain orientation.)
More recently, Nielsen and coworkers\cite{mouritsen3} have approached the
same problem in a different fashion, and placed the lipids on a random
lattice rather than on a regular lattice. The random lattice was constructed
as a network of hard disks on a plane, tethered to each other such that
the tethers cannot exceed a given maximum length. The model has been simulated
at constant pressure with a suitably adapted version of the 
dynamic-triangulation algorithm, which will be discussed in detail in section 
\ref{s42}. The interactions between lipids are taken to depend both on their
conformational state and on their distance, which introduces very
naturally a coupling between the conformation of the lipids and that
of the random lattice. If this coupling is weak, the gel-fluid transition
takes place at a higher temperature than the crystallization transition.
However, the two transitions can be brought to concur if the coupling
is chosen large enough.

Note that large density fluctuations are suppressed by construction
in a random lattice model. In order to include them, one could simply
simulate a mixture of hard disks with internal conformational 
degrees of freedom. Very simple models of this kind, where the conformational
degrees of freedom only affect the size or the shape of the disks,
have been studied by Fraser {\em et al}\cite{fraser}. They are found to
exhibit a broad spectrum of possible phase transitions.

Zhang {\em et al}\cite{zhang1} have discussed the effect of intermonolayer
coupling on the gel-fluid transition. They find that any kind of coupling
usually drives the transition to be more strongly first order. 
In other studies, the model has been generalized to incorporate 
hydrogen bonding and hydration in bilayers\cite{zhang2}. 
Moreover, it has been extended to include several molecular 
species, such that binary mixtures of lipids\cite{risbo,jorgensen}
and mixtures of cholesterol and lipids\cite{mouritsen2} could
be studied. Special attention has been given to lipid-protein 
interactions\cite{sperotto,heimburg,sabra}. Proteins in bilayers
have been reviewed recently by Mouritsen\cite{mouritsenr2}.

\section{Phenomenological models}
\label{s5}

The last class of models, which are widely used to describe amphiphilic
systems, are the phenomenological models. As opposed to all the previous
models, they totally ignore the fact that amphiphilic fluids are composed
of particles, and describe them by a few mesoscopic quantities. In
doing so, they offer the possibility to clarify the interrelations
between different behaviors on a very general level, and to study universal 
characteristica which are independent of the molecular details.

As already mentioned in the introduction, phenomenological models for
amphiphilic systems can be divided into two big classes: Ginzburg-Landau
models and random interface models.

\subsection{Ginzburg-Landau models}
\label{s51}

Ginzburg-Landau theories of amphiphiles have been reviewed at various
places\cite{gs_review,gerhardr1}, among other in chapter 11 of this book. 
Hence we shall be brief in this subsection.

The basic idea of a Ginzburg-Landau theory is to describe the system
by a set of spatially varying ``order parameter'' fields, typically
combinations of densities. One famous example is the one-order-parameter
model of Gompper and Schick\cite{gompper2}, which uses as only 
variable $\phi$ the density difference between oil and water,
distributed according to the free energy functional
\begin{equation}
\label{oop}
{\cal F}\{\phi\} = \int d \vec{r} \Big[
c (\Delta \Phi)^2 + g(\phi) (\nabla \phi)^2 + f(\phi) \Big].
\end{equation}
Here the functions $g(\phi)$ and $f(\phi)$ are defined in a suitable
way to produce the desired phase behavior (see chapter 11). The amphiphile
concentration does not appear explicitly in this model, but
it influences the form of $g(\phi)$ -- in particular, its sign. 
Other models work with two order parameters, one for the difference 
between oil and water density and one for the amphiphile density.
In addition, a vector order-parameter field sometimes accounts for the
orientional degrees of freedom of the amphiphiles\cite{gs_review}.

The equilibrium phase behavior of models of this kind has been investigated
by various methods, including Monte Carlo sampling methods\cite{kraus,holyst}, 
since these allow to account for fluctuation effects in a complete
and straightforward way. However, the vast majority of computer simulations
which have been dealing with Ginzburg-Landau models has employed them
to study nonequilibrium phenomena. In particular, the phase separation
kinetics at the presence of surfactants has attracted much interest.
It was mostly investigated in two dimensions by Langevin 
simulations\cite{laradji3,paetzold,melenkowitz,kawakatsu1,kawakatsu2}.
The differences lie in the particular form of the Ginzburg-Landau model, 
and in the dynamical system which is examined. For example,
P\"atzold and Dawson\cite{paetzold} include hydrodynamic effects and
couple the system to Navier-Stokes equations, whereas most other
groups consider a simpler relaxation dynamics, with or without conserved
order parameters.
Kawakatsu{\em et al} take special interest in situations where the
surfactant molecule is much larger than the oil or water, {\em i.e.},
a polymer. Therefore they work with a hybrid models, which treats
the surfactants as particles, and the difference of oil and water 
density by a Ginzburg-Landau field\cite{kawakatsu1}. Related models due to
the same authors represent all particle densities by continuous fields, but 
include long range interactions between the surfactants, or extend the
original hybrid model in other ways\cite{kawakatsu2}.

The phase separation process at late times $t$ is usually governed
by a law of the type $R(t) \propto t^n$, where $R(t)$ is the characteristic
domain size at time $t$, and $n$ an exponent which depends on the
universality class of the model and on the conservation laws in the
dynamics. At the presence of 
amphiphiles however, the situation is somewhat complicated by the fact that 
the amphiphiles aggregate at the interfaces and reduce the interfacial
tension during the coarsening process, {\em i.e.}, the interfacial
tension depends on the time. This leads to a pronounced slowing down
at late times. In order to quantify this effect, Laradji 
{\em et al}\cite{laradji3,laradji4} have proposed the scaling {\em ansatz} 
\begin{equation}
\label{sd}
R(t) = t^n f(\rho_s^x t) \qquad \mbox{with} \qquad x=1/n \qquad
\mbox{in two dimensions}.
\end{equation}
The function $f$ incorporates the screening effect of the surfactant, 
and $\rho_s$ is the surfactant density. The exponent $x$ can be derived from
the observation that the total interface area at late times should
be proportional to $\rho_s$. In two dimensions, this implies
$R(t) \propto 1/\rho_s$ and hence $x=1/n$. 
The scaling form (\ref{sd}) was found to describe consistently data from 
Langevin simulations of systems with conserved order parameter 
(with $n=1/3$)\cite{laradji3}, systems which evolve according to 
hydrodynamic equations (with $n=1/2$)\cite{paetzold}, and also data from 
molecular dynamics of a microscopic off-lattice model 
(with $n=1/2$)\cite{laradji1}. The data collapse has not been quite
as good in Langevin simulations which include thermal noise\cite{paetzold}.

Langevin simulations of time-dependent Ginzburg-Landau models have also been 
performed to study other dynamical aspects of amphiphilic 
systems\cite{flimmeren,kodama}. An attractive alternative approach are the 
Lattice-Boltzmann models, which take proper account of the hydrodynamics of 
the system. They have been used recently to study quenches
from a disordered phase in a lamellar phase\cite{gonnella,theissen}.

\subsection{Random interfaces}
\label{s52}

Random interface models for ternary systems share the feature with the Widom 
model and the one-order-parameter Ginzburg-Landau theory (\ref{oop}) that 
the density of amphiphiles is not allowed to fluctuate independently, 
but is entirely determined by the distribution of oil and water. However, 
in contrast to the Ginzburg-Landau approach, they concentrate 
on the amphiphilic sheets. Self-assembly of amphiphiles into monolayers
of given optimal density is premised, and the free energy of the system 
is reduced to effective free energies of its internal interfaces.
In the same spirit, random interface models for binary systems postulate
self-assembly into bilayers and introduce an effective interface
Hamiltonian to study the conformations of the bilayers.

For fluid membranes, in which neighbor relations are not maintained, the 
free energy of a membrane is often written in the 
form\cite{safran_book,gerhardr2}
\begin{equation}
\label{memb}
 \beta {\cal H} = \int d S \; \Big[ \sigma + \lambda_S H +
2 \kappa H^2 + \bar{\kappa} K \Big].
\end{equation}
Here $dS$ denotes a surface element, $H$ the local mean curvature, and $K$
the local Gaussian curvature. They are derived from the two local radii of 
curvature $R_1$ and $R_2$ {\em via} $H = (1/R_1 + 1/R_2)/2$, and
$K=1/(R_1R_2)$. The parameter $\sigma$ drives the total amount of
interface in the system and is thus in a sense related to the chemical 
potential of the amphiphiles -- it can also be interpreted as an
interfacial free energy or a negative spreading pressure. The second
term $\lambda_S$ generates a preferred radius of curvature towards one side
of the membrane, hence it breaks the symmetry of the two sides. The
parameters $\kappa$ and $\bar{\kappa}$ denote the bending rigidity and the 
saddle-splay modulus, respectively. Note that the integral over the last 
term depends solely on the topology of the interfaces according
to the Gauss-Bonnet theorem,
\begin{equation}
\label{gb}
  \int d S K = 2 \pi \chi_E, \qquad \mbox{with} \qquad \chi_E=2(c-g),
\end{equation}
where the Euler characteristic $\chi_E$ counts the number of closed
surfaces $c$ (including cavities) minus the number of handles $g$. 

How can one simulate such a system? 

A relatively simple approach suggests itself if the interfaces are known to be
almost flat. In that case, the interface position can be described
by a single-valued function $z(x,y)$, where $(x,y)$ are cartesian
coordinates on a flat parallel reference plane. The functional (\ref{memb})
can be approximated by
\begin{equation}
\label{monge}
 \beta {\cal H} = \int dx \; dy  \Big[
 \frac{\sigma}{2} (\nabla z)^2 + \frac{\kappa}{2} (\Delta z)^2 \Big]
\end{equation}
(Monge representation\cite{gs_review,safran_book}). Due to the underlying
cartesion coordinates, the discretization of this Hamiltonian is 
straightforward. Obviously, the Hamiltonian (\ref{monge}) does not really
need to be studied by Monte Carlo simulations, since it can be solved
exactly. However, things become more interesting if one considers stacks
of interacting membranes\cite{lipowsky,gompper8,netz1}, or fluid
membranes with locally varying elastic properties\cite{netz2}.

The model (\ref{monge}) is simple to study, but unfortunately not very widely 
applicable. In general, one is more interested in situations where the
interfaces are free to fold around and to assume every possible conformation.
A second possible approach is to switch over to a lattice formulation.
This has been done by a number of groups\cite{colangelo,likos,menon}.
The resulting models are very similar to the Widom or Alexander model.
As a first step, random surfaces are composed from plaquettes on a lattice. 
In models for ternary systems\cite{colangelo,likos}, these have to be closed, 
but they may be open in models for binary systems\cite{menon} (there is no 
reason why a bilayer should not end somewhere). Then, a Hamiltonian is 
introduced for these random surfaces which imposes penalties on plaquettes 
who meet at a right angle, who share a link with more than one other 
plaquette, etc. --  in the case of open surfaces also for edges and seams. 
A particularly elegant form has been used by Likos {\em et al}\cite{likos}: 
They construct the Hamiltonian as a sum over the four Minkowski functionals 
in three dimensions, {\em i.e.}, the volume, the area, the total mean 
curvature, and the Euler characteristic\cite{klaus}. If the surfaces are 
closed, the Hamiltonian can then be mapped in a straightforward way onto an 
Ising model on the dual lattice, which has various pair and multiplett 
interactions. Since the models are so similar to the lattice models of 
section \ref{s41}, their phase behavior is comparable as well. One finds 
ordered phases, three phase coexistence and a structured microemulsion in 
ternary systems, and a sponge phase in the binary system.

Thus random interfaces on lattices can be investigated rather efficiently.
On the other hand, much analytical work has concentrated on systems described 
by Hamiltonians of precisely the type (\ref{memb}), and off-lattice 
simulations of models which mimic (\ref{memb}) as closely as possible are 
clearly of interest. In order to perform such simulations, one first needs 
a method to generate the surfaces $\{ S\}$, and second a way to 
discretize the Hamiltonian (\ref{memb}) in a suitable way. 

A widely used method to generate surfaces is the dynamic 
random-triangulation algorithm\cite{kostov,billoire}, which we have 
already mentioned in section \ref{s32}. Surfaces are modeled by
triangular networks of spherical beads, linked by tethers of some
given maximum length $l_0$. The tethers define the neighbor relations on
the surface. In order to generate self-avoidance, the beads are equipped
with hard core interactions, and the maximum tether length is chosen
smaller than $\sqrt{3} \sigma$, where $\sigma$ denotes the bead 
diameter\cite{ho}. The Monte Carlo algorithm involves two different
types of steps\cite{bm_review}: Regular attempts of moving single
beads or clusters of beads in space, and attempts to change the
connectivity of the network, {\em i.e.}, to redistribute the tethers
between the chains\cite{bm_review}. The latter is done by randomly cutting 
tethers and reattaching them between the four beads which form two 
neighboring triangles\cite{kostov,billoire}, as illustrated schematically in 
Fig. \ref{fig14}. Updates of the bead positions are subject to the
constraint that the maximum tether length must not exceed $l_0$, and
connectivity updates have to comply in addition with the requirement, that 
at least three tethers are attached to one bead, and that two beads cannot be 
connected to each other by more than one tether. Otherwise, attempted
updates are accepted or rejected according to a standard Metropolis 
prescription, 

The next task is to discretize the surface integral (\ref{memb}). 
In most simulations, all terms except for the third one are dropped:
The average interfacial area is determined by the choice of the number of 
beads and the maximum tether length; including the first term would only
make sense in a ``grandcanonical'' ensemble, where beads can be added and 
removed. The last term ($ 2 \pi \chi_E$ according to eqn. (\ref{gb})) can be 
evaluated much more accurately by other methods than by an integral over 
some approximate expression for the Gaussian curvature. 
Finally, the amphiphilic layers are usually assumed to be intrinsically
symmetric, which eliminates the second term. The remaining term is the 
bending penalty on the mean curvature $H$. 
It is usually approximated by\cite{kantor}
\begin{equation}
\label{bend1}
\int dS \; 2 \kappa H^2 \approx 
\lambda \sum_{\alpha \beta} (1-\vec{n}_{\alpha} \vec{n}_{\beta})
+ 2 \pi \kappa \chi_E,
\end{equation}
where $\vec{n}_\alpha$ is the unit normal vector of triangle $\alpha$,
and the sum runs over all neighbor triangle pairs. This expression
is commonly employed in simulations of stiff fluid membranes. However, it 
has the disadvantage that the relationship between $\lambda$ and $\kappa$
is not clear. Comparing ideal spheres, for example, one obtains
$\lambda = \sqrt{3} \kappa$\cite{gompper3}, whereas a similar calculation
yields $\lambda = 2 \kappa/\sqrt{3}$ for the case of cylinders.
In order to avoid this problem, Gompper and Kroll\cite{gompper4}
have recently argued that a more appropriate discretization of
the bending free energy should be based directly on the square of
the local mean curvature:
\begin{equation}
\label{bend2}
\int dS \; 2 \kappa H^2 \approx 
\frac{\tau}{2} 
\sum_{ij} \frac{1}{\sigma_i} \Big[ 
\sum_{j(i)} \frac{\sigma_{ij}}{l_{ij}} (\vec{R}_i - \vec{R}_j) \Big]^2,
\end{equation}
with $\tau \approx \kappa$, 
where $i$ runs over all sites, $j$ over all neighbors of $i$, 
$\vec{R}_i$ denotes the position of bead $i$ and $l_{ij}$ the length of
the bond connecting $i$ with $j$. The other quantities are related to the
dual lattice, which is created from the intersections of the perpendicular 
bisectors of the bonds: $\sigma_{ij}$ is the length of the dual bond which
bisects $l_{ij}$, and $\sigma_i  = \sum_{j(i)} \sigma_{ij} l_{ij}/4$
the area of the dual cell of site $i$. Note that $\sigma_{ij}$ is 
not necessarily positive.

As yet, models for fluid membranes have mostly been used to investigate 
the conformations and shapes of single, isolated membranes, or 
vesicles\cite{ho,boal,gompper3,gompper4,bernd1,gompper5,ipsen}.
In vesicles, a pressure increment $p$ between the vesicle's
interior and exterior is often introduced as an additional relevant
variable. An impressive variety of different shapes has been 
found, including branched polymer-like conformations, inflated vesicles,
dumbbell shaped vesicles, 
and even stomatocytes. Fig. \ref{fig15} shows some typical configuration
snapshots, and Fig. \ref{fig16} the phase diagram for vesicles of size 
$N=247$, as calculated by Gompper and Kroll\cite{gompper5}.

The collapsed polymer-like state is characterized by specific scaling laws, 
{\em e.g.}, the average volume of the vesicle is propertional
to the number of beads, $\langle V \rangle \propto N$.
Other scaling laws apply in other regions of the phase diagram.
For the case of zero pressure increment, $p=0$, and moderate bending 
rigidity, a scaling 
{\em ansatz} of the form
\begin{equation}
\label{sc}
\langle V \rangle = N^{3/2} \theta(\sqrt{A}/\xi)
\end{equation}
has been proposed\cite{gompper5,ipsen}, where $\sqrt{A} = \sqrt{N}$ is 
the area of the vesicle, and $\xi \propto \exp(4 \pi /3 \kappa)$ its 
persistence length. If the scaling law (\ref{sc}) holds, simulation data for 
different vesicle sizes $N$ should collapse on one curve, if
$\langle V \rangle N^{-3/2}$ is plotted versus 
$\sqrt{N} \exp(-4 \pi \kappa/3)$.
This is indeed the case\cite{gompper5,ipsen}. 
Scaling laws are extremely sensitive tools which allow to study various
phenomena very accurately. For example, it has been predicted\cite{peliti2}
that the bending rigidity $\kappa$ is softened by fluctuations on large 
length scales $l$
\begin{equation}
\kappa(l) = \kappa - \frac{\alpha}{4 \pi} \ln (l/a_0),
\end{equation}
where $a_0$ is a microscopic cutoff length, and $\alpha$ a universal constant.
By a careful scaling analysis of volume fluctuations in the limit of large
bending rigidity, Gompper and Kroll have been able to verify this 
prediction\cite{gompper4}.

A number of recent studies consider more complex systems, such as freezing 
vesicles\cite{gompper6} (freezing can be induced by reducing the tether 
length), or mixed membranes which contain more than one 
component\cite{kumar,koyama}. The possibility that a membrane may
break up and form pores has also been considered\cite{shillcock}.

In all of these investigations, the topology of the simulated object was 
kept fixed (spherical). Current work is devoted to study systems with
variable topology, where vesicles can fuse and break up. Such models will
probably be very useful to study bicontinuous states, and the evolution
of them into states of isolated vesicles and 
droplets\cite{bernd2,gompper7}.

\subsection{Conclusions}

We have attempted to give an overview over the wide spectrum of 
topics which are currently investigated in amphiphilic systems, and 
over the multitude of simulation methods and simulation models which
have been used to explore them. Amphiphilic systems have been studied
on length scales ranging from a few Angstrom to micrometers, and
over a similarly wide range of time scales. Hence a whole hierarchy
of models have been developed, each of which covering a different
length scale, and devised to address a different type of problem. 
We have also attempted to give a feeling for the many interesting
questions which are still open and/or under current investigation.
Amphiphilic systems turn out to be an immensely rich playground
for researchers, which has something to offer for almost everybody:
for the materials scientist, the physical chemist, the biologist,
the condensed matter physicist, and the hard core theoretician in
statistical physics. 

\section*{Acknowledgements}

It is a pleasure to thank Michael Schick, Christoph Stadler,
and Harald Lange for enjoyable collaboration, and Kurt Binder
for fruitful discussions and encouragement. I have benefitted
from stimulating interactions with Mark Matsen, Ralph Blossey,
Marcus M\"uller, Andreas Werner, Frank Haas and Nigel Wilding.
Thanks also go to Gregor Huber for carefully reading this
manuscript, and to the Deutsche Forschungsgemeinschaft for
financial support through the Heisenbergprogramm.

\clearpage

\newpage

\noindent
Self-assembled structures in amphiphilic systems:
Micellar structures (a) and (b) exist in aqueous solution as
well as in ternary oil/water/amphiphile mixtures. In the
latter case, they are swollen by the oil on the hydrophobic
(tail) side. Monolayers (c) separate water from oil domains
in ternary systems. Lipids in water tend to form bilayers (d)
rather than micelles, since their hydrophobic
block (two chains) is so compact and bulky compared to the 
head group, that they cannot pack into a sphere very 
well\cite{israelachvili}. At small concentrations, bilayers often 
close up to form vesicles (e). Some surfactants also
form cylindrical (wormlike) micelles (not shown).

\newpage

\begin{figure}
\fig{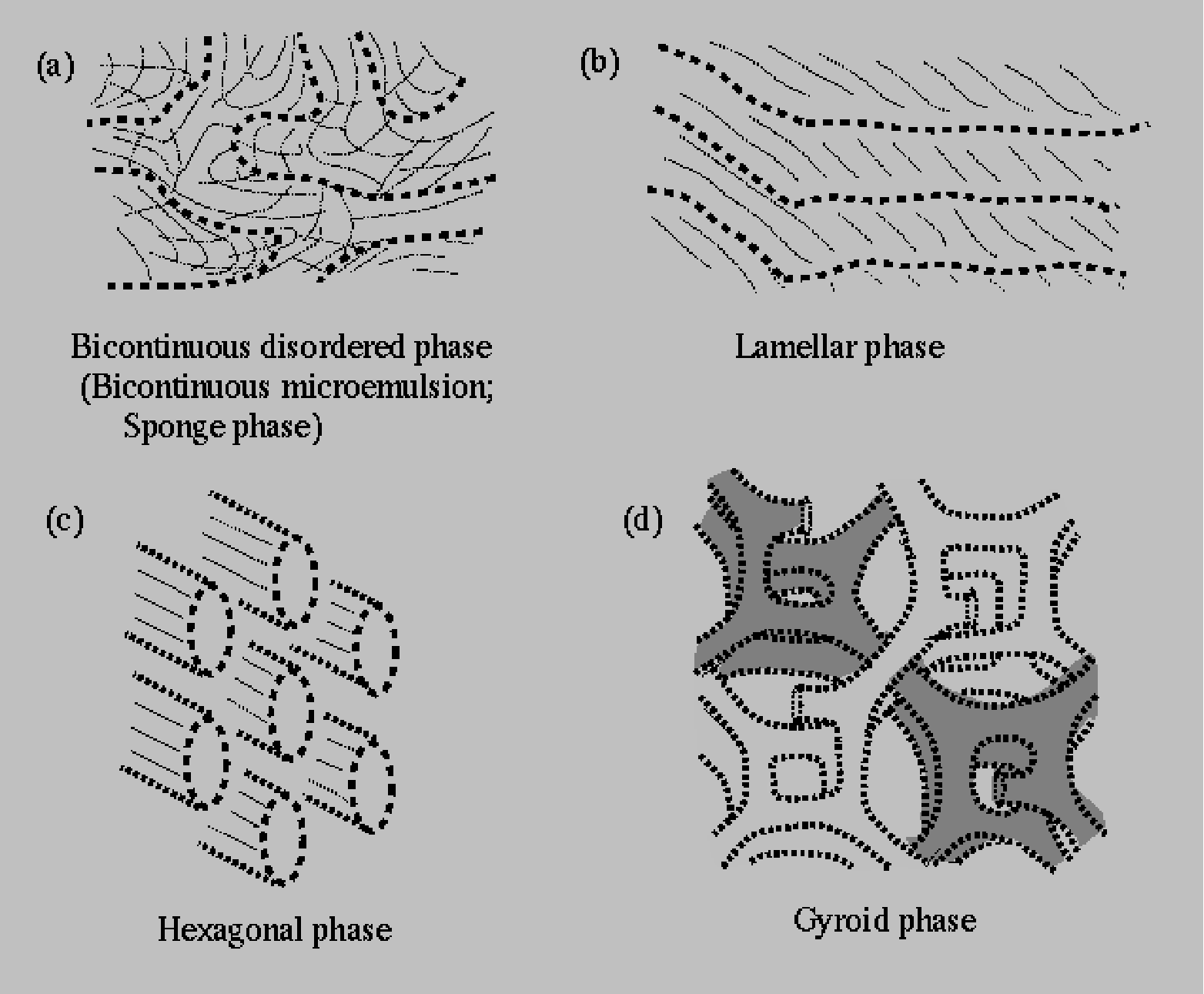}{130}{110}
\caption{\label{fig2}}
\end{figure}
\noindent
Selected structured phases in amphiphilic systems:
Again, these phases are present in both ternary oil/water/amphiphile
mixtures and binary systems of water and amphiphile. In the first
case, the dashed lines represent monolayers which separate oil from water 
domains, in the second case, they represent in (a) and (b) bilayers,
in (c) and (d) surfaces of amphiphile aggregates. Specifically, we show
a macroscopically isotropic, but microscopically structured
bicontinuous phase, called sponge phase ($L_3$ structure) 
in the binary mixture\cite{sponge} and bicontinuous microemulsion in 
the ternary mixture\cite{microemulsion} (a);
the lamellar phase $L_{\alpha}$(b); the hexagonal phase $H_1$(c); 
and the gyroid phase $G$ (d). The different shading in (d) 
distinguishes between the two constituting networks of the gyroid.
See text for more explanation.


\begin{figure}
(a) \fig{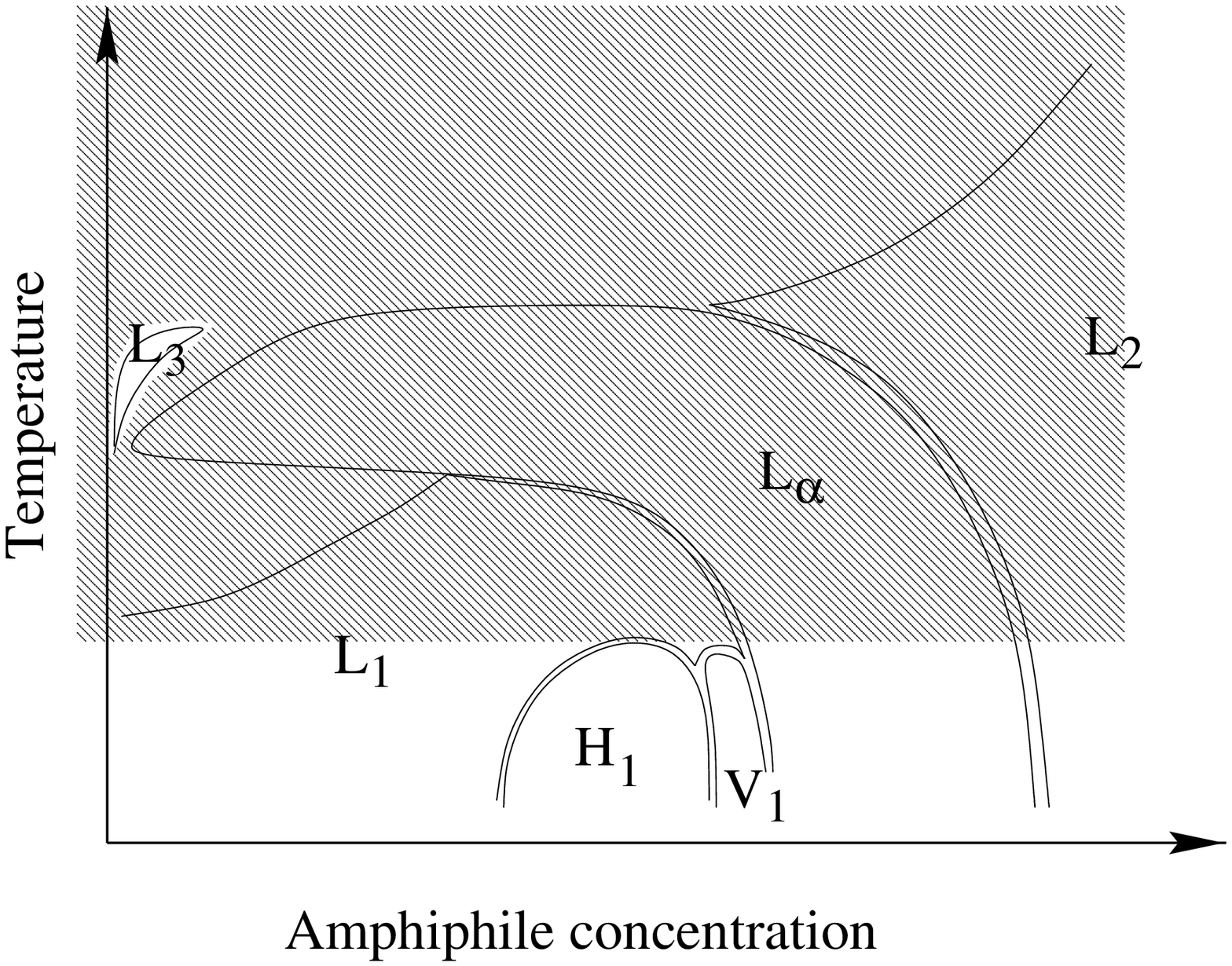}{90}{75}
(b) \fig{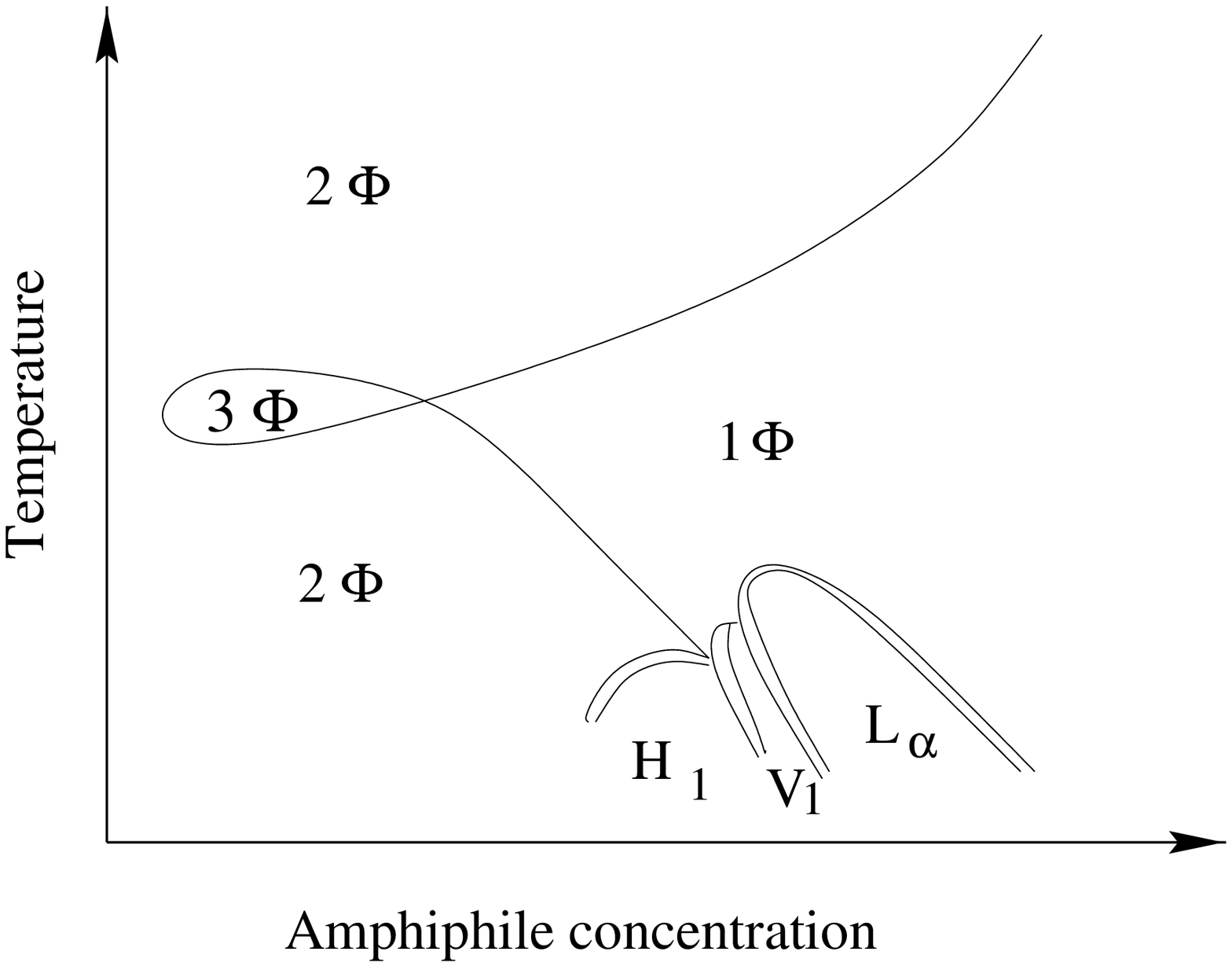}{90}{75}
\caption{\label{fig3}}
\end{figure}


\newpage

\noindent
Fig. 3 \\
\noindent
Schematic phase diagrams of amphiphilic systems with good anionic amphiphiles
(a) for a binary water/amphiphile mixture 
(adapted from Strey {\em et al} 1990\cite{strey1}),
and (b) for a ternary mixture at an oil/water ratio of 1:1
(adapted from Kahlweit {\em et al} 1986\cite{strey2}).
Here $L_{\alpha}$ denotes the lamellar phase, $H_1$ the hexagonal
phase and $V_1$ a phase with cubic symmetry, presumably a gyroid phase.
In the binary system (a), $L_3$ represents the sponge phase
and $L_1, L_2$ two other isotropic liquid phases. 
Shading indicates regions of two phase coexistence. In the
ternary system, the symbol $1 \Phi$ labels a single phase region,
$2 \Phi$ a region of two phase coexistence between an oil-rich and
a water-rich phase, and $3 \Phi$ a region of three phase coexistence
with an additional amphiphile rich ``middle phase''. In systems with
strong amphiphiles, the coexisting middle phase is usually a
structured microemulsion (cf. Fig. \ref{fig2}).

\newpage

\begin{figure}
\fig{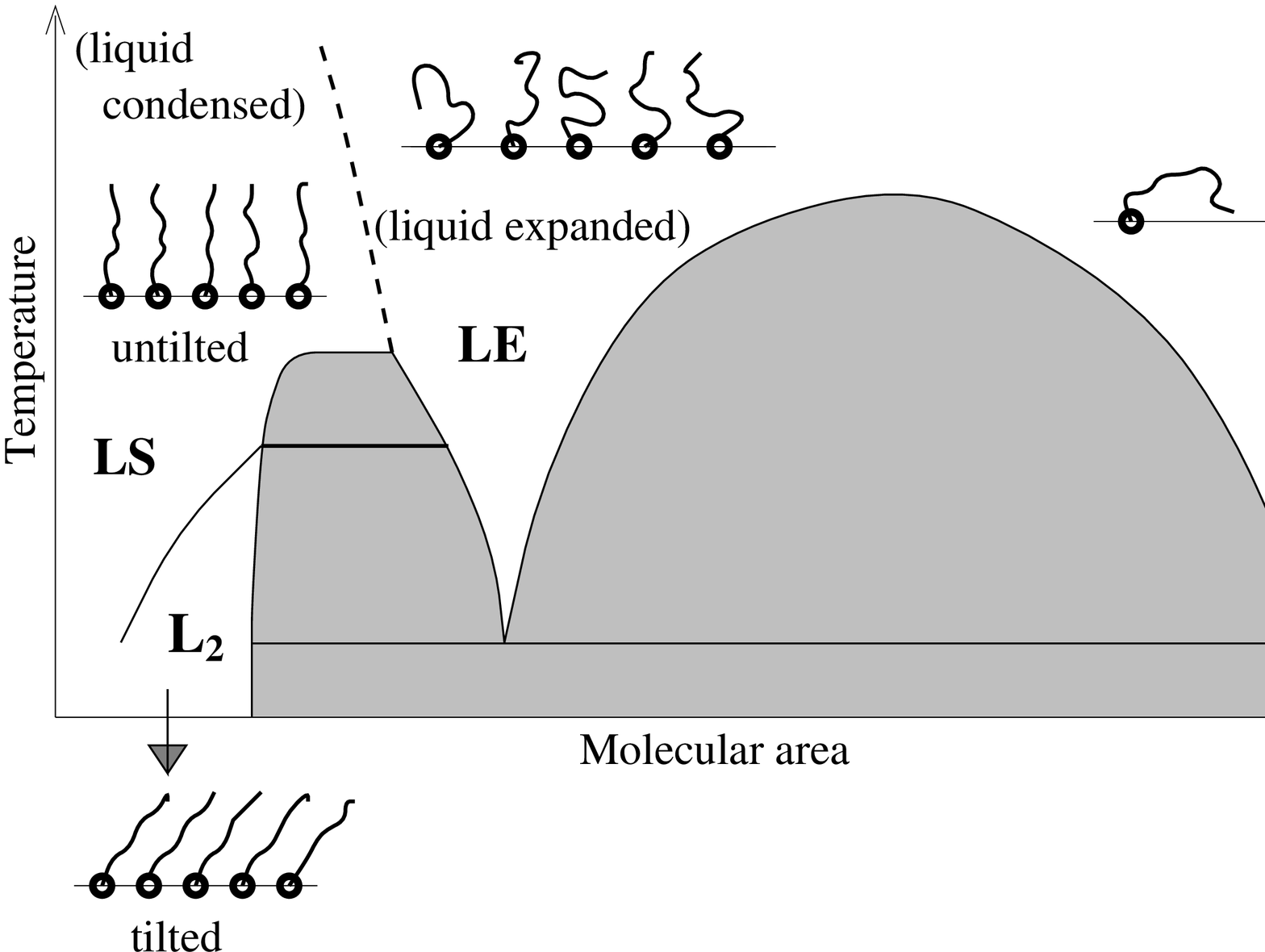}{120}{100}
\caption{\label{fig4}}
\end{figure}
\noindent
Phase diagram of Langmuir monolayers at low and intermediate surface
coverage (schematic). Not shown are the various phases on the condensed
side at high surface coverage.

\newpage

\begin{figure}
\fig{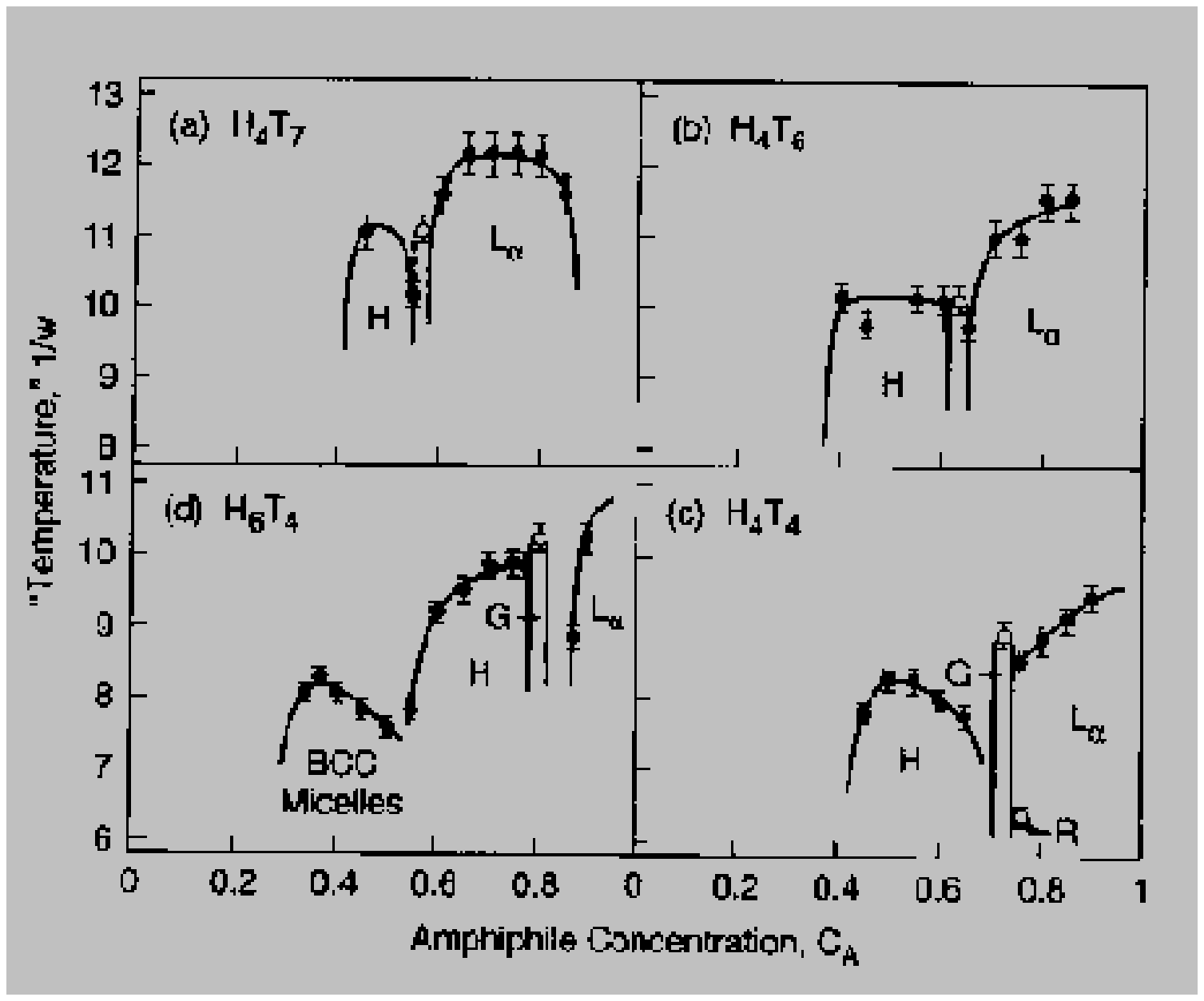}{120}{100}
\caption{\label{fig5}}
\end{figure}
\noindent
Phase diagrams of binary amphiphile/water systems in the Larson model.
The amphiphile structures are 
(a) $H_4 T_7$, (b) $H_4 T_6$, (c) $H_4 T_4$ and (d) $H_6 T_4$.
$H$ denotes the hexagonal phase, $L_{\alpha}$ the lamellar phase,
$G$ the gyroid phase, $BCC$ a phase with spherical micelles in 
body-centered cubic arrangement, and $R$ a rhombohedral-like mesh phase. 
From Larson 1996\cite{larson4}.

\newpage

\begin{figure}
\fig{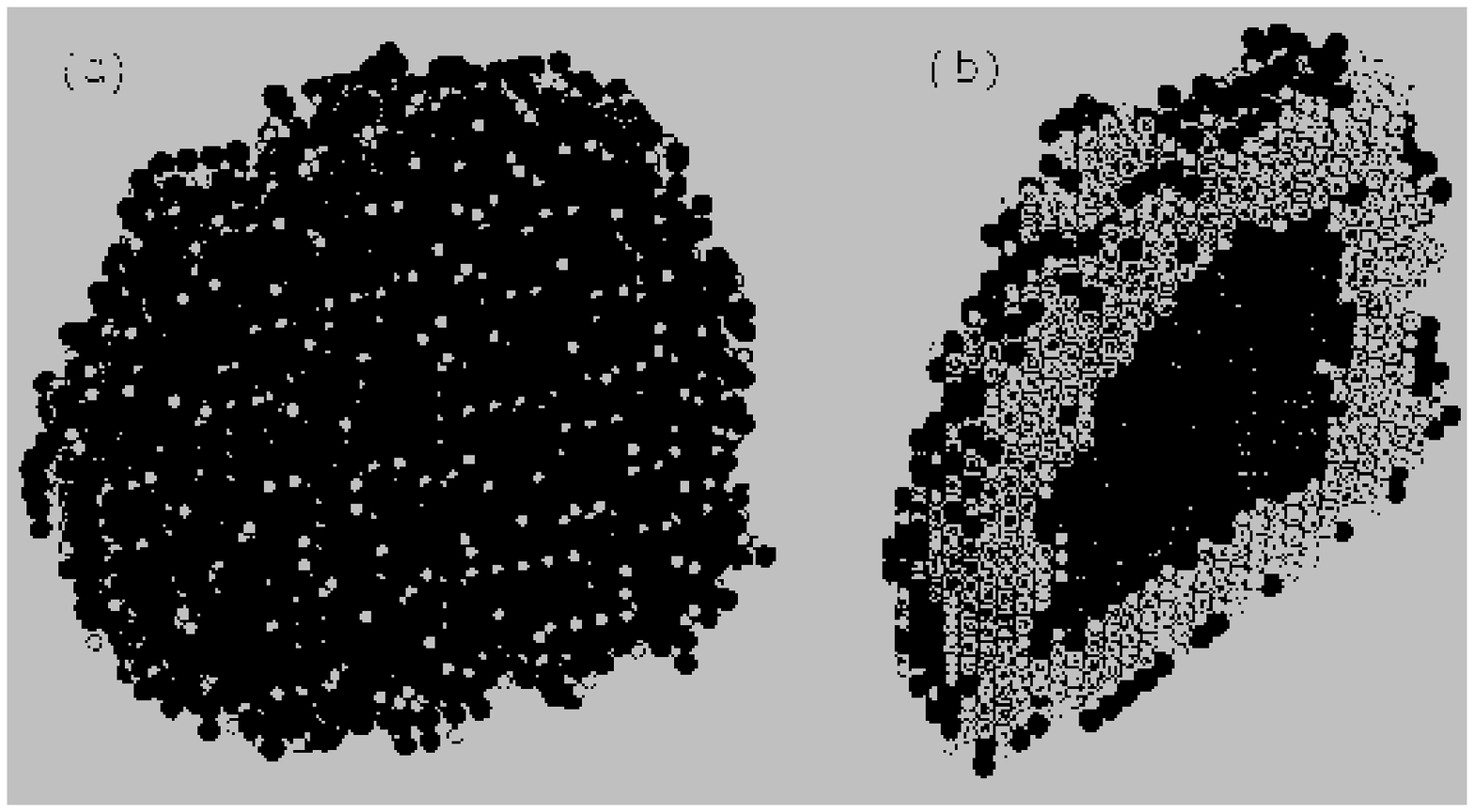}{120}{80}
\caption{\label{fig6}}
\end{figure}
\noindent
Configuration snapshot of a spontaneously formed vesicle from double-tailed
amphiphiles in the Larson model. (a) entire vesicle (b) vesicle cut in half
in order to show its inner side. Black circles represent head particles (+1),
grey circles tail particles (-1), white circles the neutral connecting
particles (0). 
From Bernardes 1996\cite{bernardes5}.

\newpage

\begin{figure}
\fig{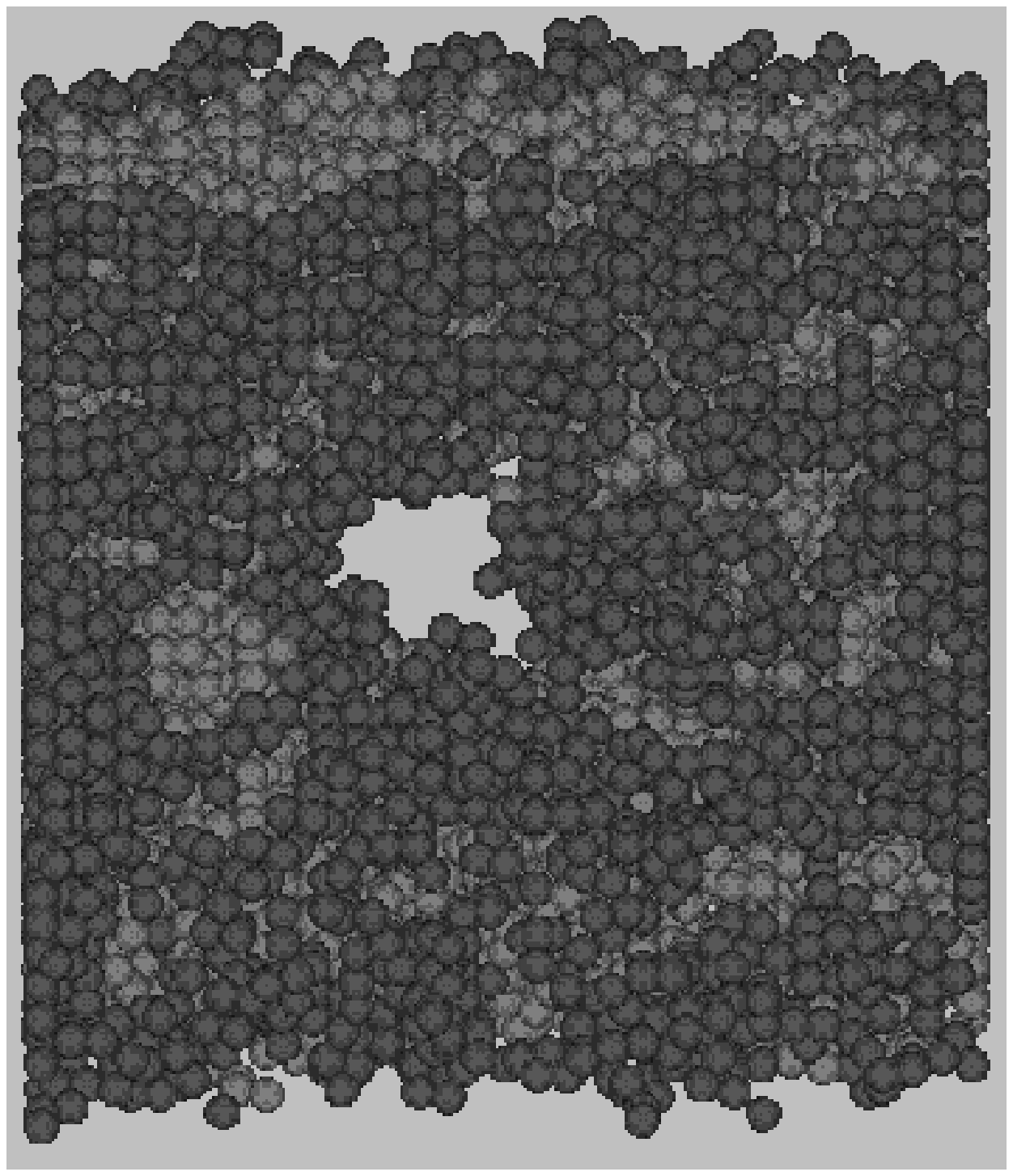}{100}{90}
\caption{\label{fig7}}
\end{figure}
\noindent
Snapshot of a bilayer conformation with a pore in the bond-fluctuation
model. The dark spheres represent head particles, the light spheres
tail particles. Around the pore, the amphiphiles rearrange so as
to shield the bilayer interior from the solvent.
From M\"uller and Schick 1996\cite{marcus}.

\newpage

\begin{figure}
\fig{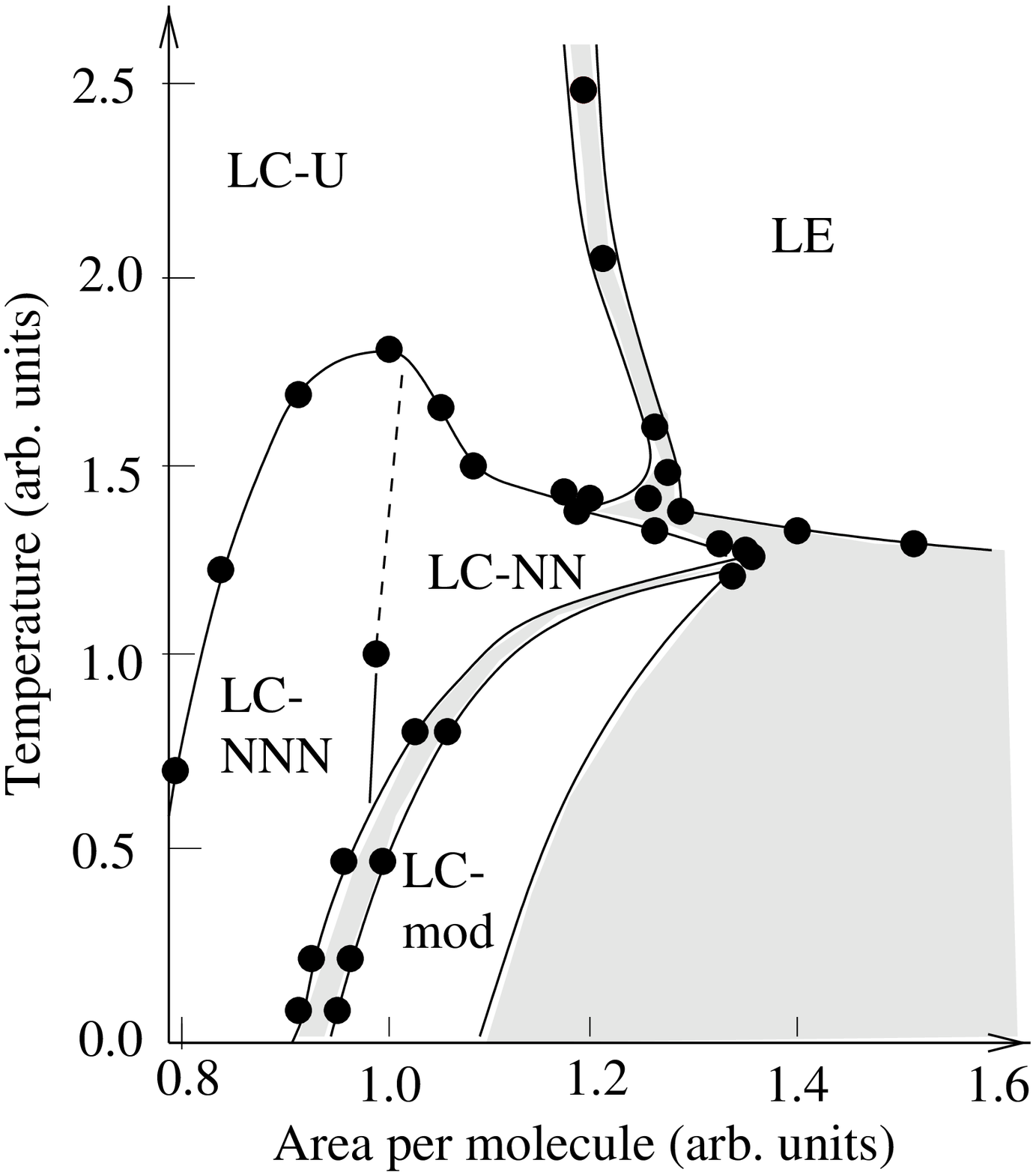}{100}{110}
\caption{\label{fig8}}
\end{figure}
\noindent
Phase diagram of a Langmuir monolayer in a model of grafted stiff 
Lennard-Jones chains. LE denotes a disordered expanded phase, 
LC-U a condensed phase with untilted chains, LC-NN and LC-NNN 
condensed phases with collective tilt towards nearest neighbors and 
next nearest neighbors, respectively, and LC-mod a phase which
has a superstructure and an intermediate direction of tilt.
From Stadler and Schmid 1998\cite{christoph2}.

\newpage

\begin{figure}
(a) \fig{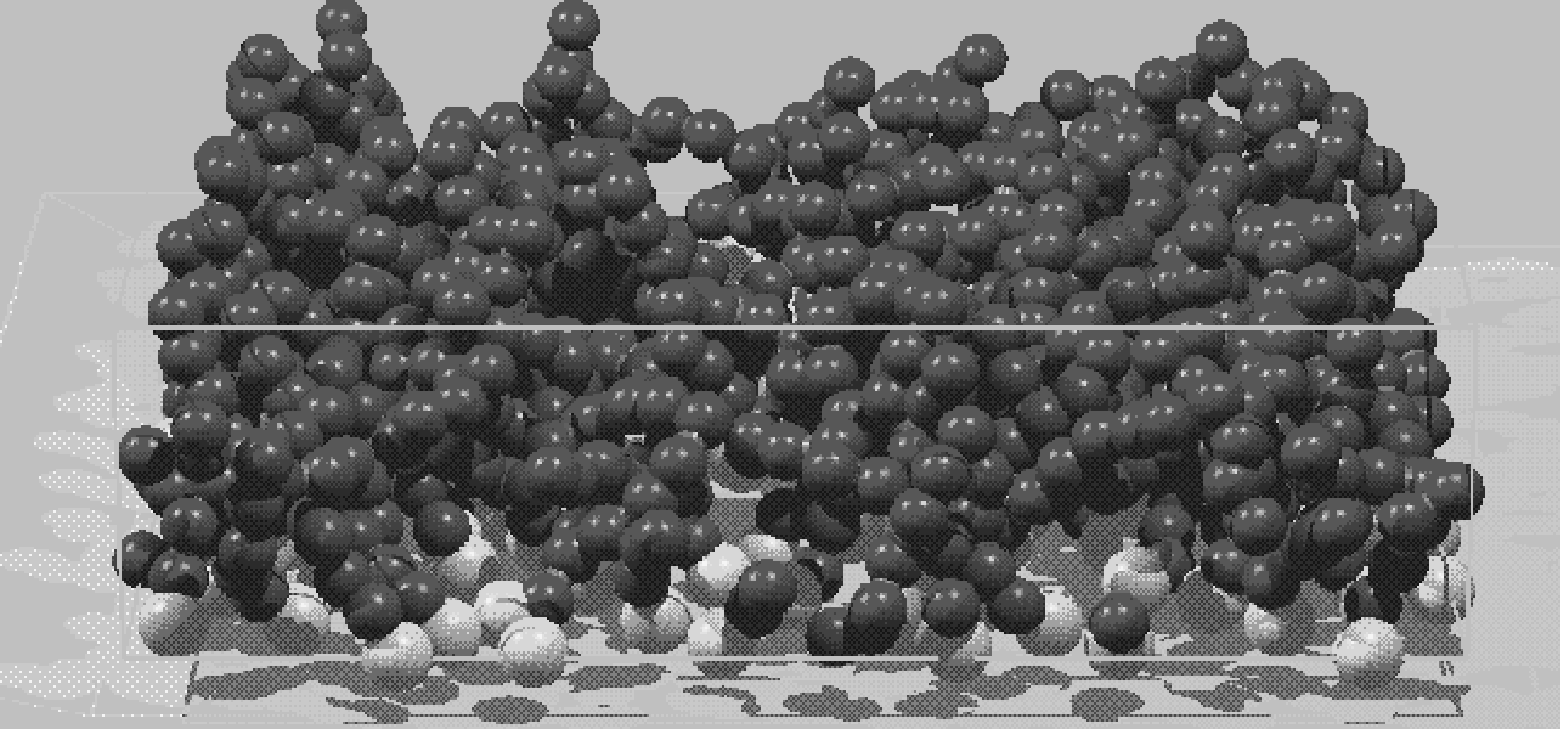}{100}{50}
(b) \fig{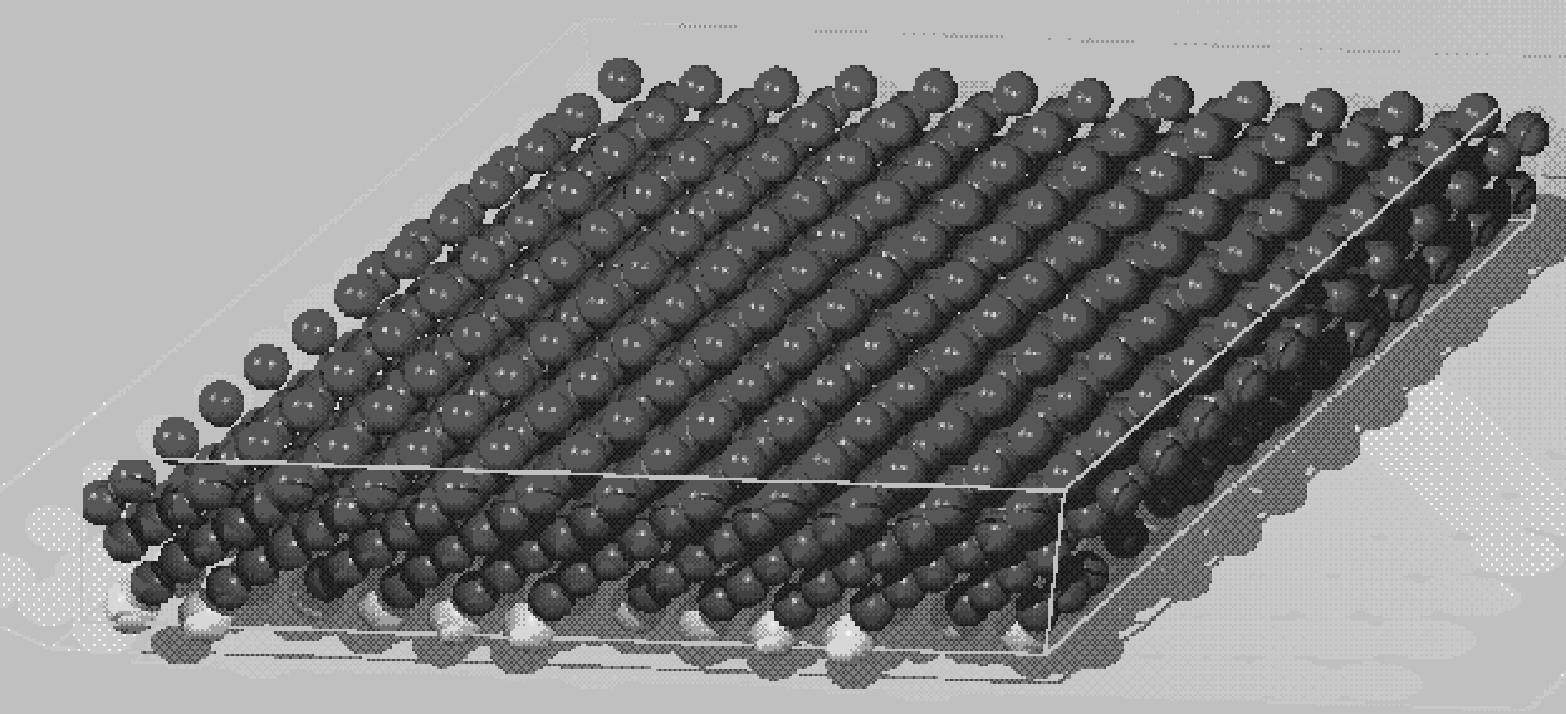}{100}{65}
\caption{\label{fig9}}
\end{figure}
\noindent
Configuration snapshots of the monolayer in Fig. \ref{fig8}
(a) in the disordered expanded phase LE, 
(b) in the condensed modulated phase LC-mod.
From Stadler and Schmid 1998\cite{christoph2}.

\newpage

\begin{figure}
\fig{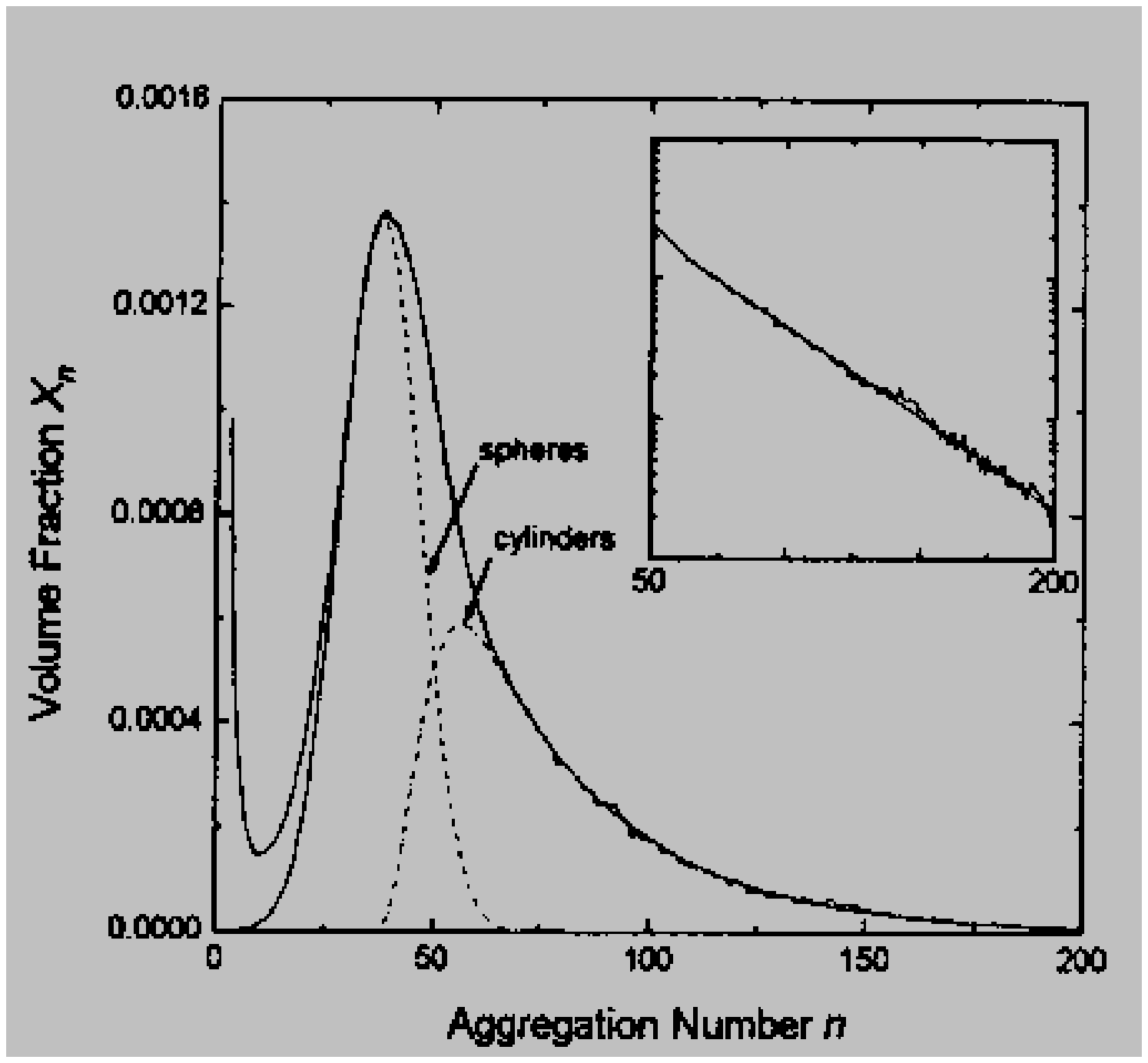}{120}{100}
\caption{\label{fig10}}
\end{figure}
\noindent
Micelle size distribution for $H_2T_2$ surfactants within the Larson model.
The dashed lines show fits to the expected form for spherical micelles
(main peak) and cylindrical micelles (tail). Inset shows the tail of
the distribution on a semi-logarithmic plot to demonstrate the
exponential decay predicted for the cylindrical micelles.
From Nelson, Rutledge and Hatton 1997\cite{nelson}.

\newpage

\begin{figure}
\fig{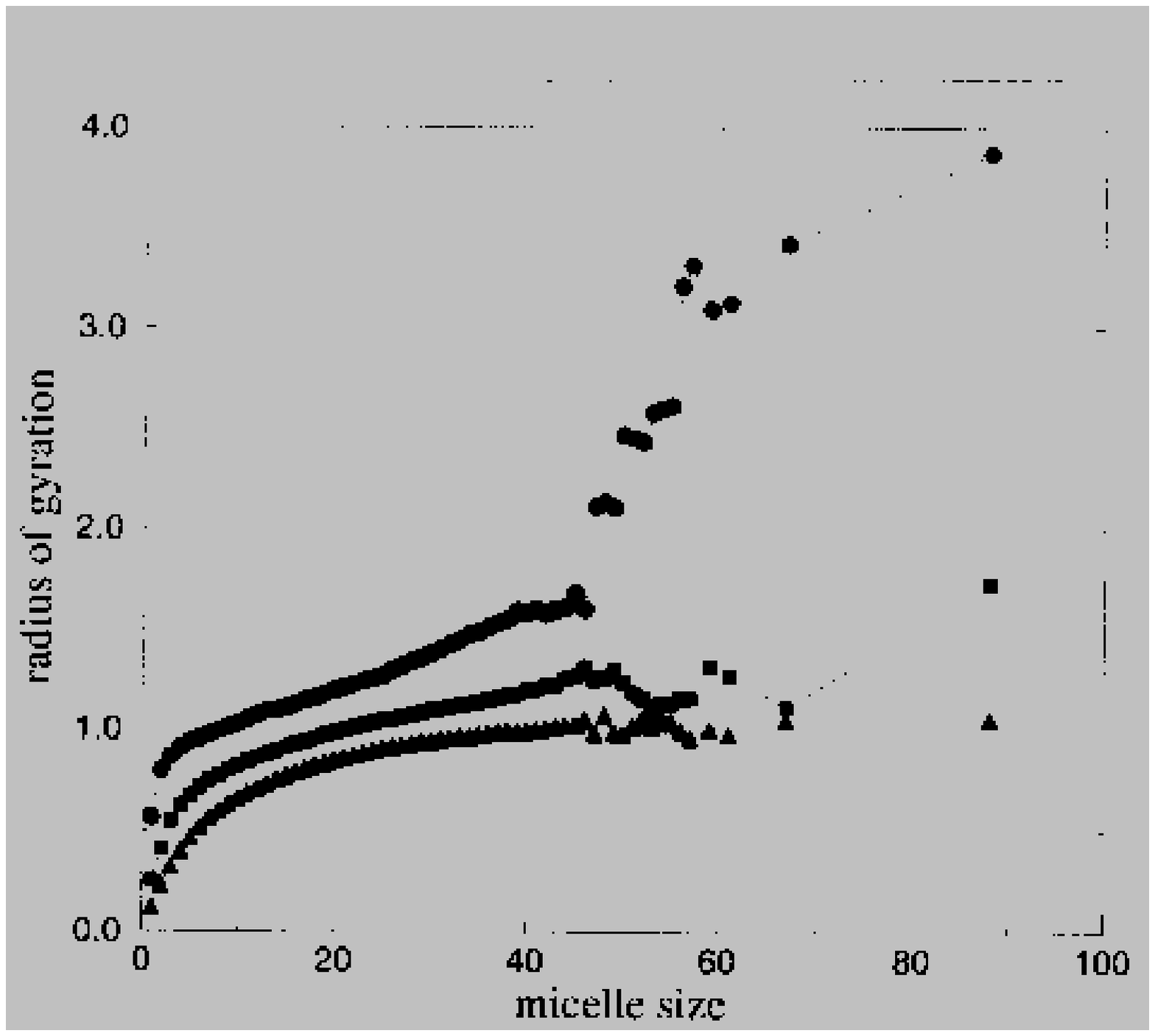}{120}{100}
\caption{\label{fig11}}
\end{figure}
\noindent
Eigenvalues of the radius of gyration tensor (dots: largest, squares: middle, 
triangles:smallest) of micelles vs. aggregation number $N$ in an
off-lattice model of $H_2T_2$ surfactants. The micelle size distribution
for this particular system has a peak at $N\approx 28$.
From Viduna, Milchev and Binder 1998\cite{viduna}.

\newpage

\begin{figure}
\fig{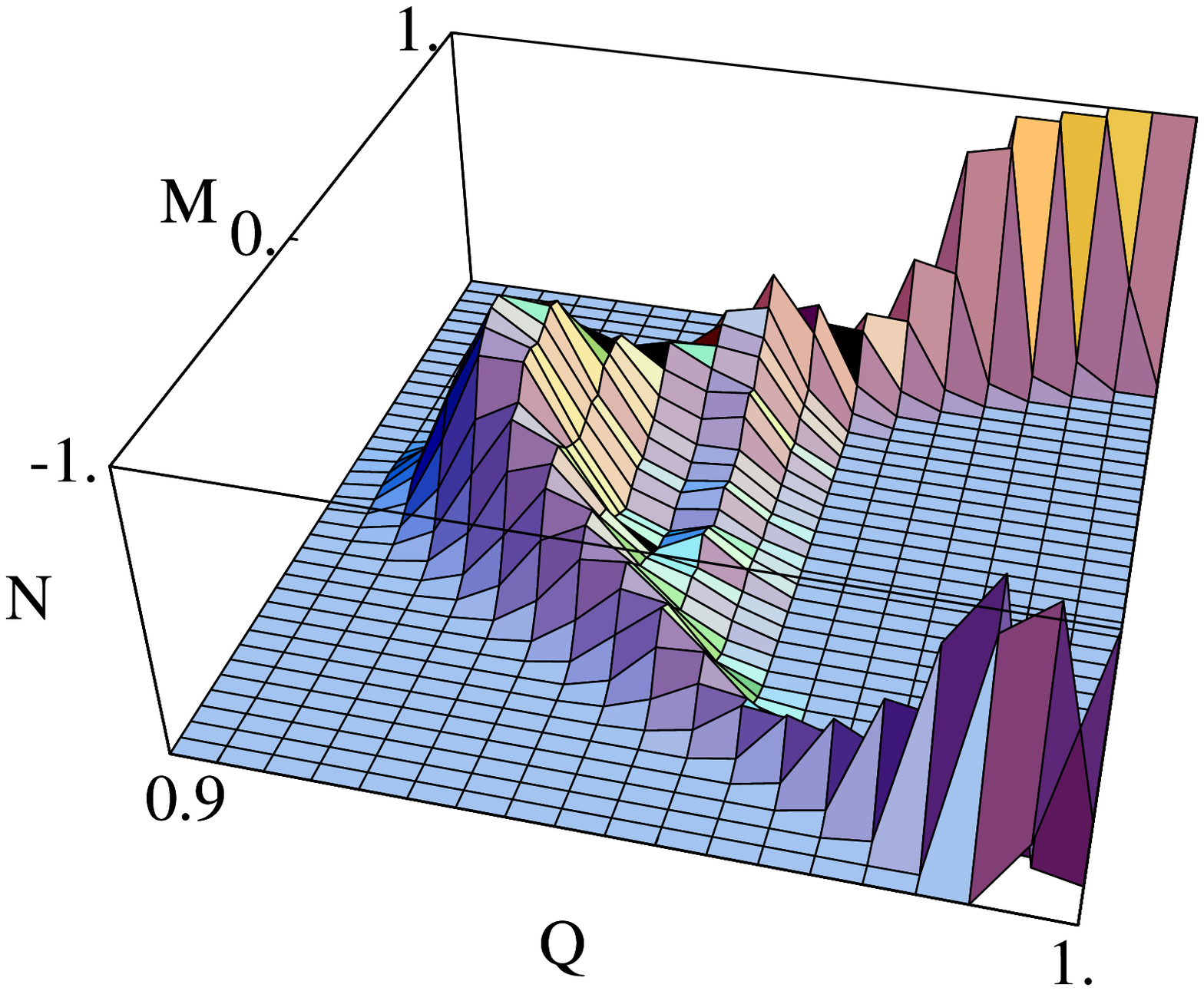}{100}{130}
\caption{\label{fig12}}
\end{figure}
\noindent
Histogram of $Q$ and $M$ in the three component model at three
phase coexistence. Parameters are $C=H=0,K=0.5,\Delta=-7,L=-5$ in units
of $J$, and temperature $k_B T/J=2.78$. System size is $12\times 12\times24$.
See text for further explanation.
From Schmid and Schick 1994\cite{friederike4}.

\newpage

\begin{figure}
\fig{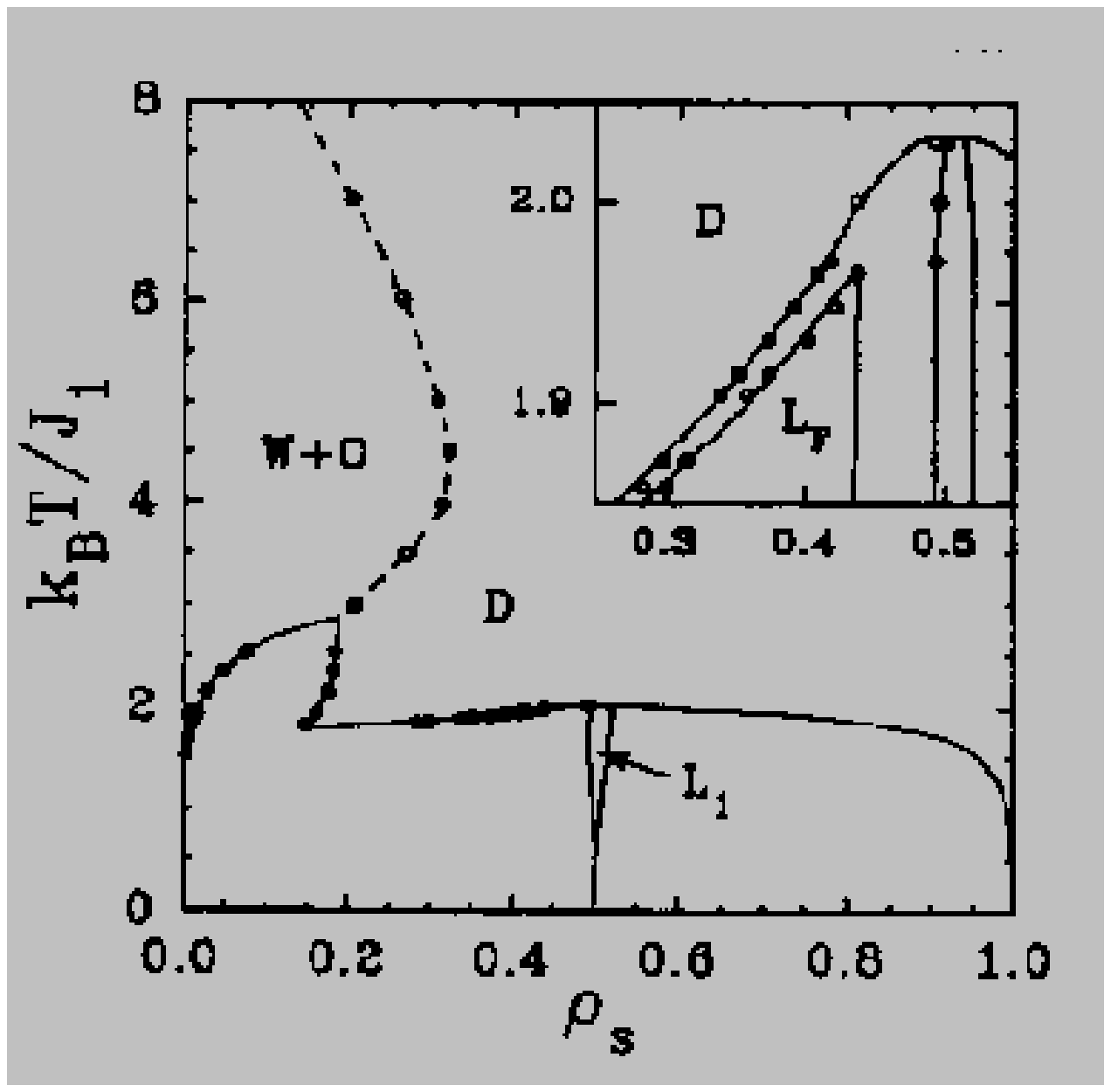}{100}{90}
\caption{\label{fig13}}
\end{figure}
\noindent
Phase diagram of a vector lattice model for a balanced ternary amphiphilic
system in the temperature vs. surfactant concentration plane. 
W+O denotes region of coexistence between oil- and water-rich
phases, D a disordered phase, L${}_1$ an ordered phase which consists
of alternating oil, amphiphile, water and again amphiphile sheets,
and L${}_F$ an incommensurate lamellar phase (not present in mean field
calculations). The data points are based on simulations at various system
sizes on an fcc lattice.
From Matsen and Sullivan 1994\cite{mark1}.
Copyright 1994 by the American Physical Society.

\newpage

\begin{figure}
\fig{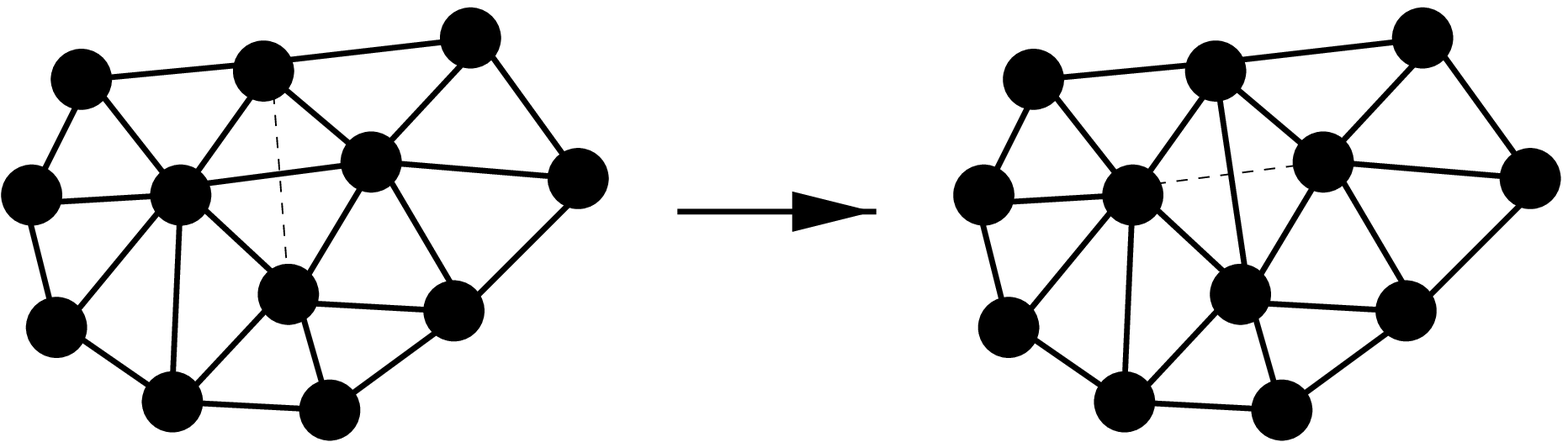}{100}{35}
\caption{\label{fig14}}
\end{figure}
\noindent
Schematic sketch of a connectivity update in the dynamic triangulation
algorithm.

\newpage

\begin{figure}
\fig{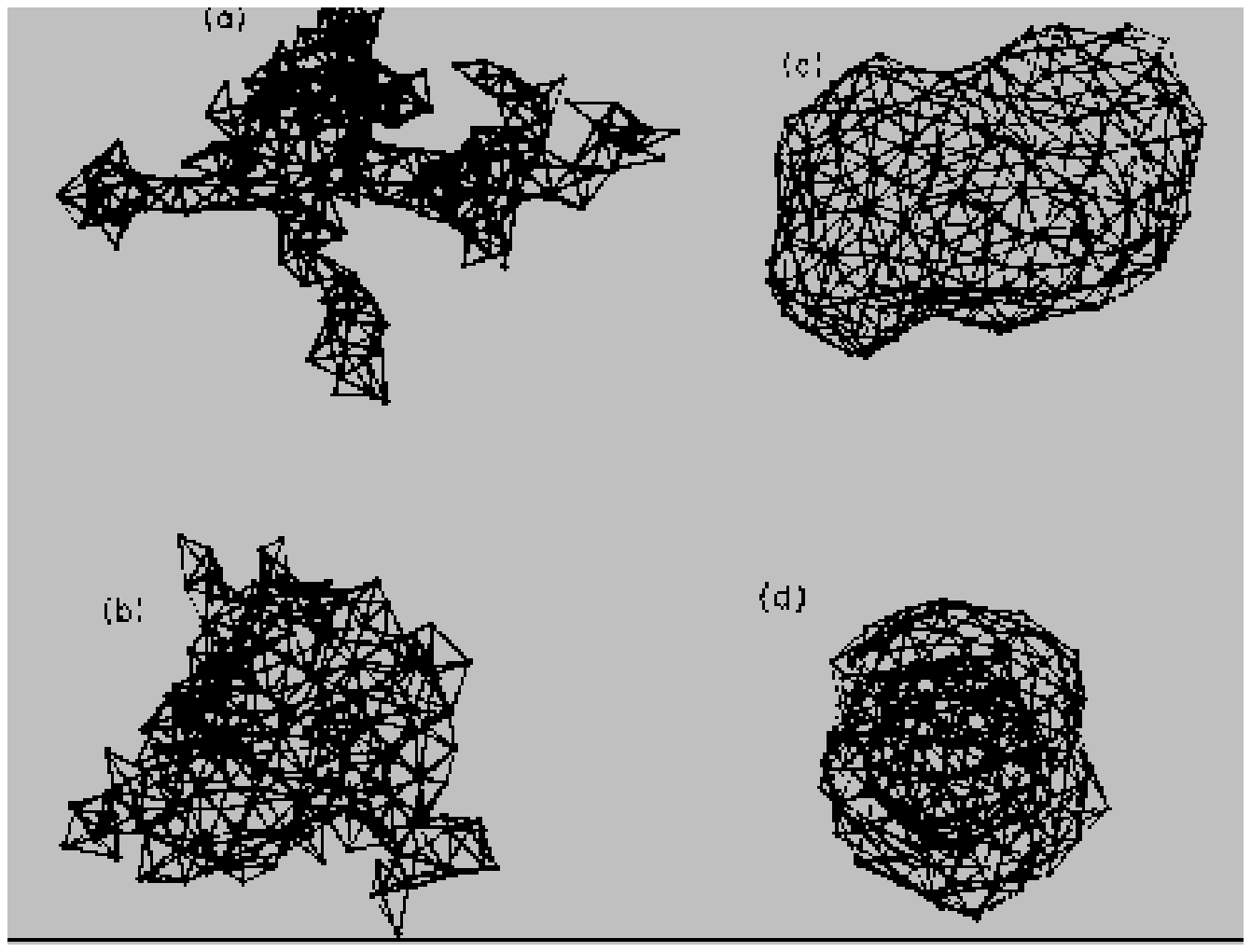}{120}{100}
\caption{\label{fig15}}
\end{figure}
\noindent
Conformations of fluid vesicles for different values of the bending rigidity
and pressure increment.
(a) branched polymer
(b) inflated vesicle
(c) prolate vesicle
(d) stomatocyte.
From Gompper and Kroll 1995\cite{gompper5}.
Copyright 1995 by the American Physical Society.

\newpage

\begin{figure}
\fig{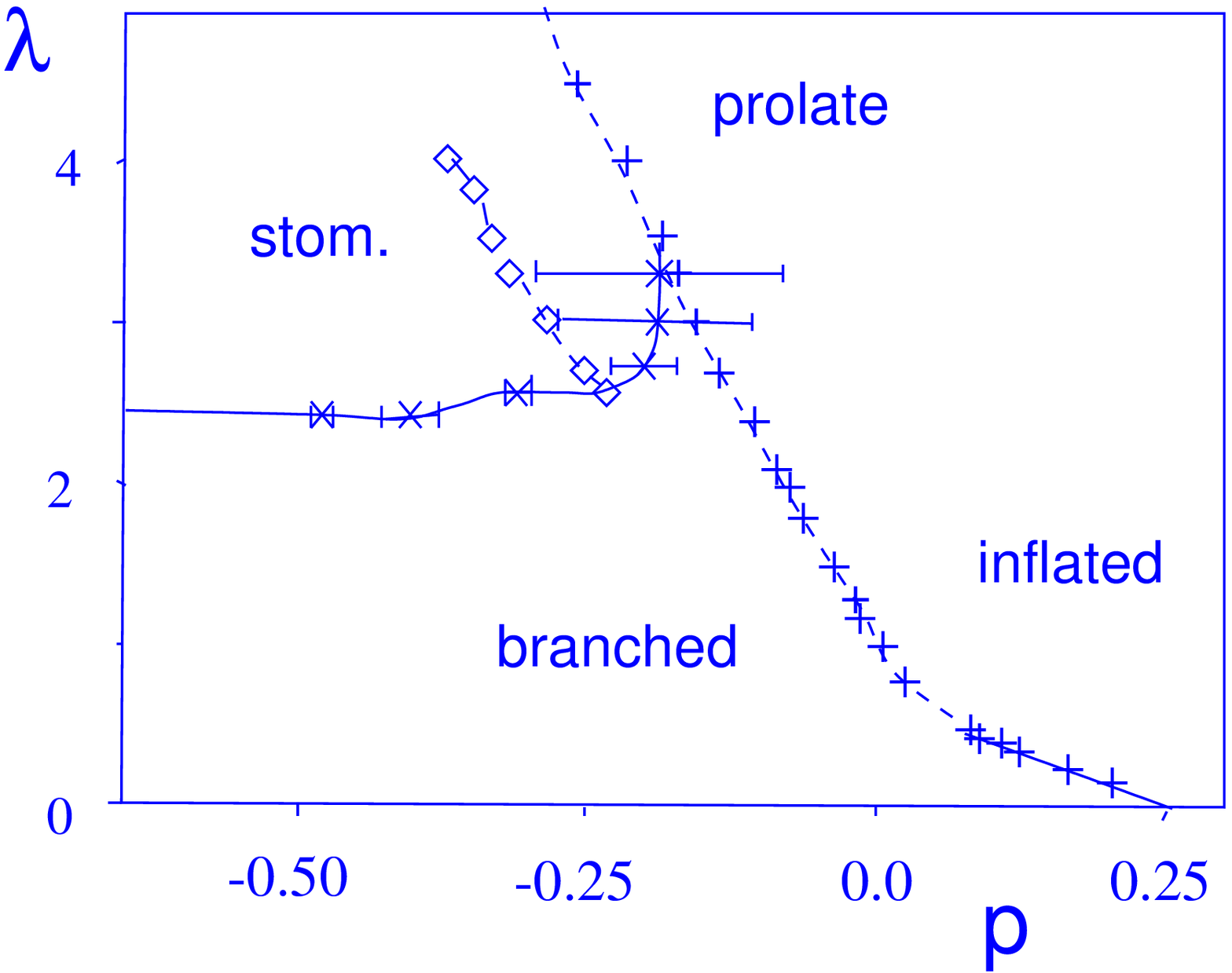}{100}{80}
\caption{\label{fig16}}
\end{figure}
\noindent
Phase diagram of fluid vesicles as a function of pressure increment $p$ 
and bending rigidity $\lambda$. Solid lines denote first order transitions,
dotted lines compressibility maxima. The transition between the
prolate vesicles and the stomatocytes shows strong hysteresis effects,
as indicated by the error bars. Dashed line (squares) indicates a
``transition'' from metastable prolate to metastable diskshaped vesicles.
From Gompper and Kroll 1995\cite{gompper5}.


\begin{thebibliography}{99}
\bibitem{fn1}
  Of course, these numbers are constantly increasing due to the fast
  progress in computer technology.
\end{thebibliography}
\end{document}